\newcommand{\rom}[1]{\uppercase\expandafter{\romannumeral #1\relax}}
\documentclass[onecolumn, 12pt]{IEEEtran}
\usepackage{amsmath}
\usepackage{amsthm}
\usepackage{amssymb}
\usepackage{cite}
\usepackage{amsfonts}
\usepackage{enumerate}
\usepackage{subfigure}
\usepackage{multirow}
\usepackage{verbatim}
\usepackage{url}
\usepackage{color}
\usepackage{graphicx}
\usepackage{hhline}
\usepackage[dvips,dvipdfm]{geometry}
\usepackage{setspace}
\DeclareMathOperator*{\argmin}{arg\,min}

\begin{document}
\title{\huge Optimal Training for Residual Self-Interference for Full-Duplex One-way Relays}
\vspace{-10mm}
\author{Xiaofeng Li, Cihan Tepedelenlio\u{g}lu \IEEEmembership{Member, IEEE},\footnote{Xiaofeng Li and Cihan Tepedelenlio\u{g}lu are with the Department of Electrical, Computer and Energy Engineering, Arizona State University, Tempe, AZ, 85287, USA (email: xiaofen2,cihan@asu.edu).}\\
 and Habib \c{S}enol \IEEEmembership{Member, IEEE}\footnote{Habib \c{S}enol is with Department of Computer Engineering, Faculty of Engineering and Natural Sciences, Kadir Has University, Istanbul 34083, Turkey (e-mail: hsenol@khas.edu.tr).}}
\maketitle
\vspace{-20mm}
\begin{IEEEkeywords}
Full-duplex relays, residual self-interference, maximum likelihood estimation, optimal training sequence, Toeplitz matrix, frequency-selective channels, multiple relays
\end{IEEEkeywords}
\begin{abstract}
Channel estimation and optimal training sequence design for full-duplex one-way relays are investigated. We propose a training scheme to estimate the residual self-interference (RSI) channel and the channels between nodes simultaneously. A maximum likelihood estimator is implemented with Broyden-Fletcher-Goldfarb-Shanno (BFGS) algorithm. In the presence of RSI, the overall source-to-destination channel becomes an inter-symbol-interference (ISI) channel. With the help of estimates of the RSI channel, the destination is able to cancel the ISI through equalization. We derive and analyze the Cramer-Rao bound (CRB) in closed-form by using the asymptotic properties of Toeplitz matrices. The optimal training sequence is obtained by minimizing the CRB. Extensions for the fundamental one-way relay model to the frequency-selective fading channels and the multiple relays case are also considered. For the former, we propose a training scheme to estimate the overall channel, and for the latter the CRB and the optimal number of relays are derived when the distance between the source and the destination is fixed. Simulations using LTE parameters corroborate our theoretical results.
\end{abstract}

\section{Introduction}\label{sec:intro}
Due to the growing demands on wireless bandwidth, the need for high spectral efficiency has become more urgent. In-band full-duplex (FD) relays, which transmit and receive simultaneously in the same frequency band, offer a viable solution since they theoretically have the ability to double the spectral efficiency, compared to half duplex relays. In practice, however, the relay receives strong self-interference in FD mode, which is challenging to overcome. Recently, self-interference cancellation techniques have been developed with great promise \cite{Heino15full,hong2014sic5g,sabharwal2014band}. With these techniques, the self-interference can be canceled by estimating the self-interference channels\cite{ma2009new,masmoudi2015mlsic,Koohian2015estifull} in FD mode, or be suppressed with null-space methods in MIMO systems\cite{riihonen2011mitigation}. However, despite these advances, residual self-interference (RSI) still exists after the self-interference cancellation\cite{sabharwal2014band,riihonen2011mitigation,Duarte12full,riihonen2010residual}. Therefore, accurate channel estimation in the presence of RSI is required at the destination to further improve the system performance by canceling RSI.

Several works analyze the system performance in the presence of RSI with different criteria such as interference power, outage probability, and bit error rate (BER)\cite{riihonen2010residual,Kim2012full,jimenez2014per,jimenez2014power}. Reference \cite{Kim2012full} investigates the outage probability of an amplify-and-forward (AF) FD relay. Reference \cite{riihonen2010residual} takes advantages of multiple antennas to suppress the RSI power using a null-space pre-coding matrix. References \cite{jimenez2014per} and \cite{jimenez2014power} analyze the diversity and the capacity of FD relays in the presence of RSI. However, these works do not consider the cancellation of the RSI. Some of the works assume perfect channel state information (CSI) \cite{Kim2012full,jimenez2014per} while others assume imperfect CSI \cite{riihonen2010residual}, but they do not mention training schemes for FD systems. As shown by the results of these works, the system performance suffers from the RSI since it is still quite high as compared to the desired received signal, and does not yield good performance when treated as noise. In \cite{Kim2012full}, the authors assume that the RSI power is as high as 10 dB than the desired signal power and also results in inter-symbol interference (ISI) in AF relays. References \cite{sabharwal2014band}, \cite{Duarte12full}, and \cite{riihonen2010residual} report that the power of RSI may be as high as 30 dB than the noise floor even after applying self-interference cancellation. As the estimation error in analog cancellation\cite{sabharwal2014band}, RSI cannot be further estimated at the relay, which motivates incorporating the RSI into the end-to-end channel model, and then estimating it, and removing it at the destination.

In addition to the one-way relay system, in-band FD can also be applied in two-way relays in which two sources exchange their messages through the relay\cite{cheng2013twrfull,Tabataba12PA,kim2013effects,zheng2015joint,li2016fulltwr}. The achievable rates of FD relays and half duplex relays are compared in \cite{cheng2013twrfull}. References \cite{Tabataba12PA,kim2013effects} consider the channel estimation error as an additive noise terms and analyze the effect of it without showing specific training schemes. In contrast to them, we propose training schemes for the FD relay system in this paper. Reference \cite{zheng2015joint} considers an amplify-and-forward FD two-way relay system in the presence of RSI with multiple antennas. However, \cite{zheng2015joint} assumes the RSI can be eliminated by designing precoding matrices where the product of the precoding matrix and the RSI channel matrix equals to zero, which is impractical. The channel estimation problem for RSI in FD two-way relays is addressed in \cite{li2016fulltwr}. The RSI channel, which is the loop self-interference channel from the relay receive antenna to its transmit antenna after self-interference cancellation, and the channels between nodes are estimated simultaneously at the sources, and the asymptotic behavior of the Fisher information is analyzed.

The AF FD relay system can be also combined with MIMO\cite{Mohammadi16,Lemos15}. In the MIMO case, the signal distortion caused by hardware impediments like the limited dynamic-range of non-ideal amplifiers, oscillators, ADCs, and DACs has to be considered\cite{Cirik16,Day12,Day12bi,Tagh15,Tagh16} when modeling the RSI channel. However, with sufficient passive self-interference suppression and analog cancellation in RF \cite{sabharwal2014band}, the distortion can be ignored\cite{Duarte12full}, which leads to a Gaussian model for the RSI channel\cite{zheng2015joint,Ngo2014mimofull,Xiong16}. The Gaussian model still works when the distortion has to be considered. That is because the distortion is well modeled as additive Gaussian noise terms which can be incorporated into the system noise variance\cite{Cirik16}.

From the RSI channel estimation point of view, reference \cite{Xiong16} proposes two methods for the RSI channel in an AF relay system where the destination is equipped with massive MIMO. In the first method the authors consider the case where the RSI channel is estimated by the relay itself, and in the second method the base station estimates the RSI channel. However, in their system model, the RSI is incorporated into the noise term and the ISI caused by the RSI is treated as noise. A conference version of this manuscript\cite{li2015maximum} proposes a maximum likelihood (ML) estimator for the RSI channel and derives the Cramer-Rao bounds (CRBs). In this work, we further analyze the CRB by using asymptotic properties of Toeplitz matrices and design the optimal training sequences. We also extend our training method to the case in which the channels in the relay system are frequency-selective and the case of multi-relay systems, which were not considered in \cite{li2015maximum}. Reference \cite{li2016fulltwr} investigates the channel estimation problem in FD two-way relays which is fundamentally different than the one-way relay setting considered herein. We consider the one-way relay case where only the source sends training and the relay forwards in contrast to  \cite{li2016fulltwr} which considers a specific training scheme in two-way relays in where the relay sends its own training sequence. In this paper we eliminate one parameter of interest and integrate out a nuisance parameter, making the objective function depend on only a single parameter. In contrast, the ML estimator in \cite{li2016fulltwr} is with respect to multiple parameters due to the differences in the system model. From the analysis point of view, we are able to obtain a closed-form expression for the CRB and find the optimal training sequence accordingly, which is a step further than the Fisher information analysis in \cite{li2016fulltwr}. Moreover, we have extensions to the frequency-selective and multi-relay cases. The closed-form expression for the CRB is also extended in the multi-relay case to determinate the optimum number of relays.

In this work, we consider an AF FD one-way relay system with the relay working in FD mode. The RSI in the relay propagates to the destination, creating an end-to-end ISI channel. We further cancel the RSI at the destination by estimating the RSI channel and applying equalization. Different from studies which focus on canceling the self-interference at the relay itself, we reduce the complexity burden at the relay and aim to cancel the interference at the destination. Thus, the destination performs the estimation and cancellation operations. To estimate the RSI channel as well as the channels between nodes, an ML estimator is proposed, which is formulated by maximizing the log-likelihood function through a quasi-Newton method. The CRBs are derived in closed-form. By using asymptotic properties for Toeplitz matrices, we are able to find the corresponding optimal training sequences through minimizing the CRB. We also extend our training method and analysis for the basic relay system to the case when the channels between nodes are frequency-selective and the case of multiple relays. \textcolor{blue}{The main contribution of this paper is that the effective end-to-end channel of the FD relay system in the presence of RSI is investigated and an ML estimation of the effective channel including the RSI channel is proposed. The CRB and optimal training sequences are also analyzed in closed-form. We will briefly explain how to address practical issues such as the synchronization of the relay received signal and the RSI signal, and the possibility of clipping of the received signal due to the relay gain in Section \ref{sec:sysmodel}.}
 
The rest of the paper is organized as follows: Section \ref{sec:sysmodel} describes the system model of an AF FD one-way relay system. Section \ref{sec:channelest} proposes the training scheme and the ML estimator. The CRBs are derived and analyzed in Section \ref{sec:CRB}, and the optimal and the approximately optimal training sequences are also discussed. Section \ref{sec:freq} and \ref{sec:multi} extend our training method to frequency-selective channel and multi-relay case. Section \ref{sec:simulation} shows the numerical results and Section \ref{sec:conclusion} concludes the paper.

\section{System Model}\label{sec:sysmodel}
\begin{figure}
    \centering
    \includegraphics[width=8cm]{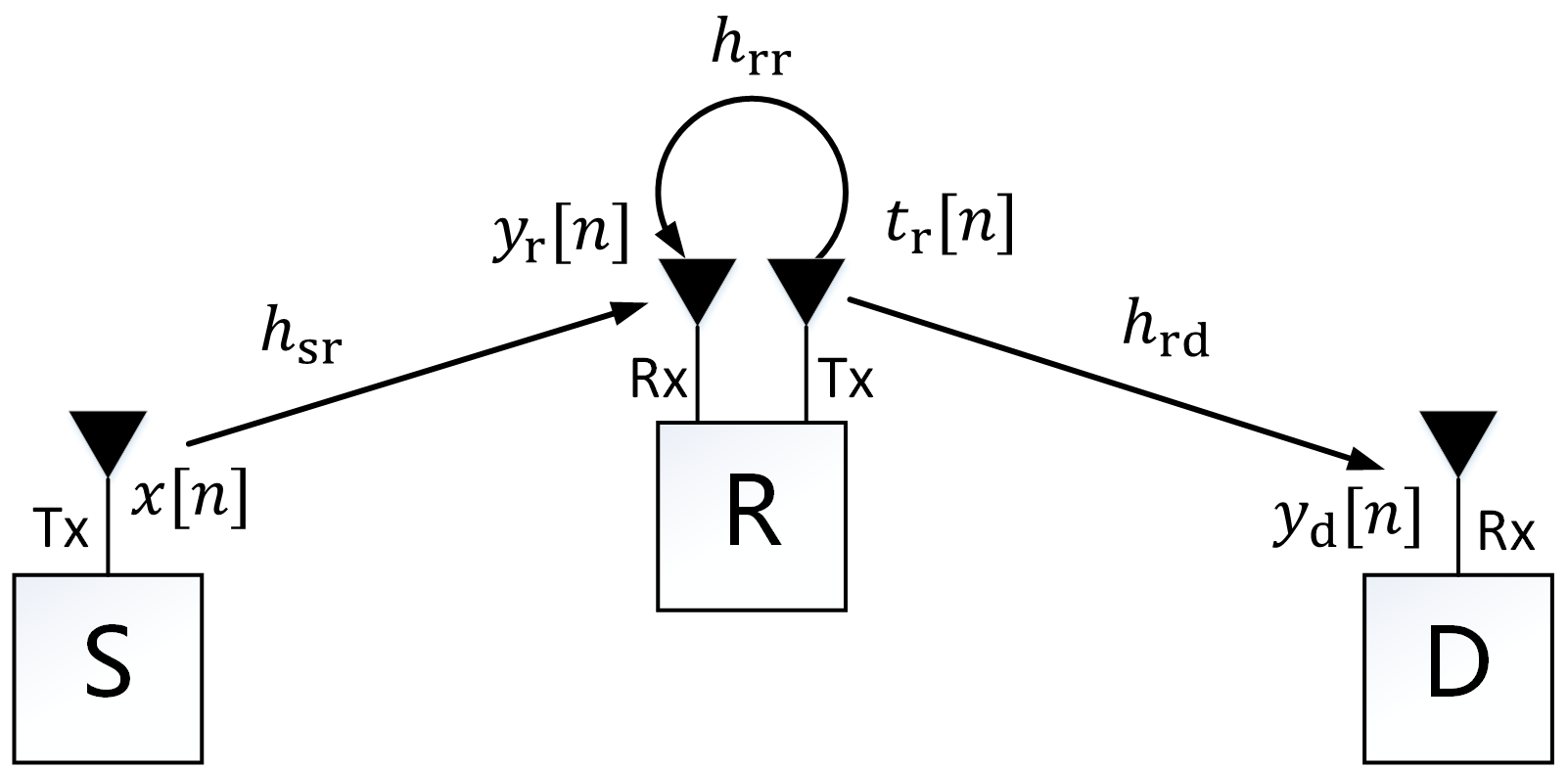}
    \vspace{-4mm}
    \caption{A full-duplex one-way relay system.}
    \label{fig.sysmodel}
    \vspace{-8mm}
\end{figure}

We consider a system consisting of a source, a relay, and a destination, without any direct link between the source and the destination, as shown in Figure \ref{fig.sysmodel}. Amplify-and-forward relay protocol is adopted. The relay uses two antennas, one receiving the current symbol while the other one amplifying and forwarding the previously received symbol, to operate in FD mode \cite{Kim2012full}. The channel coefficients between the source and the relay, and the relay and the destination are $h_{\rm sr}$ and $h_{\rm rd}$ respectively. The two channels between nodes are assumed to be flat fading modeled by independent complex Gaussian random variable with zero means and variances $\sigma_{\rm sr}^2$, $\sigma_{\rm rd}^2$, respectively. A separate pre-stage is assumed to gather the information of the self-interference channel to perform analog and digital cancellation methods in the next transmission stage\cite{sabharwal2014band,Xiong16}. During the transmission, the self-interference is reduced in RF with analog cancellation methods until the residual self-interference power falls in the ADC dynamic range, and then is further suppressed by digital methods. However, despite these suppression methods, the RSI is still present at the destination. We consider the RSI as the residual (error) through either analog cancellation only or analog-plus-digital cancellation. The non-zero residual is an unavoidable result of the self-interference cancellation. Even though the LoS component is largely canceled, the RSI power is still not small enough to be treated as noise, and is often higher than the desired signal power \cite{sabharwal2014band,Duarte12full}. Moreover, the RSI makes the overall end-to-end channel an ISI channel at the AF relay, even when the channels on all links are flat fading. Thus, estimating the RSI at the destination is needed for equalizers to alleviate the ISI at the destination receiver. 

Since the line of sight (LoS) component of the self-interference varies very slowly, it can be much reduced by the RF cancellation, which means the reflected multi-path component dominates the RSI in the transmission stage\cite{Ngo2014mimofull,Xiong16}. Under the assumption that the LoS component is reduced and the bulk of the interference is from scattering components, the RSI channel $h_{\rm rr}$ can be modeled as a complex Gaussian random variable with zero mean and variance $\sigma_{\rm rr}^2$ \cite{Ngo2014mimofull}. Moreover, the RSI channel is also assumed to be frequency-flat and time-invariant within blocks. The Gaussian model is also used by \cite{zheng2015joint} and \cite{Xiong16} for mathematical tractability. In our system model, we assume that $h_{\rm rr}$ is time-invariant and flat fading in one transmission block and varies from block to block. The Gaussian assumption is used in the simulations to generate realizations of the channels for multiple blocks but not in the derivation of our training scheme and analysis. Different channel models such as Rician model for the self-interference channel before active cancellation \cite{Duarte12full} can also be adopted.

We assume the processing delay for the relay to forward its received symbols is $\tau_0$ which is a integer multiple of the symbol duration $T_s$, i.e $\tau_0=mT_s$,\ $m=1,2,\cdots$. The processing delay $\tau_0$ is a deterministic system parameter and can be known at the system design level once the hardware and the self-interference cancellation approaches are chosen. We introduce an artificial additional processing delay to make $\tau_0$ an integer multiple of the symbol duration, which can be implemented by analog delay filters \cite{delayfilter}. With the synchronized signal, the resulting discrete-time equivalent channel model is sparse with zero coefficients and can simplify the analysis. We use $\tau_0=mT_s$ in the derivation of the ML estimator and CRB analysis but we show that it is possible to update our ML estimator and CRB derivation and analysis with an additional unknown parameter $\tau_0$ (which means no synchronization at the relay) in Appendix \rom{2}. Let the transmitted signal at the relay be $t_{\rm r}[n]=\alpha y_{\rm r}[n-m]$ where $\alpha$ is a real and positive power scaling factor. The factor controls the stability of the system through scaling the relay transmit power. We will discuss the choices of it later in this section. At the destination, the received symbol at the $n$th time interval is
\begin{align}\label{eqn:rec_sig1}
y_{\rm d}[n]&=h_{\rm rd}t[n]+n_{\rm d}[n]=h_{\rm rd}(\alpha y_{\rm r}[n-m])+n_{\rm d}[n]\nonumber\\
&=\sum_{k=1}^{\infty}h\theta^{k-1}x[n-km]+\sum_{k=1}^{\infty}d\theta^{k-1}n_{\rm r}[n-km]+n_{\rm d}[n]\ \ \ n=0,1,\cdots
\end{align}
where $y_{\rm r}[n]=h_{\rm sr}x[n]+\alpha h_{\rm rr}y_{\rm r}[n-m]+n_{\rm r}[n]$ is the $n$th received symbol at the relay. $x[n]$ is the transmitted signal of the source and satisfies $E\big[|x[n]|^2]=P_{\rm s}$ where $P_{\rm s}$ is the transmit power of the source and incorporates the path loss. For brevity, we define $d:=\alpha h_{\rm rd}$, $h:=\alpha h_{\rm sr}h_{\rm rd}$ and $\theta:=\alpha h_{\rm rr}$. Noise terms $n_{\rm r}[n]$ and $n_{\rm d}[n]$ are complex Gaussian with zero mean and variance $\sigma_{\rm r}^2$ and $\sigma_{\rm d}^2$ respectively. If there is no RSI, the effective end-to-end channel $h$ is the overall channel for the system. However, the self-interference link $\theta$ forms a feedback at the relay, which makes the overall channel a single pole infinite impulse response (IIR) channel and causes ISI. Additionally, the effective noise at the destination is colored with correlations that depend on the pole. The overall IIR channel has channel taps $[\underbrace{h,0,\cdots,0}_{m \mathrm{\ terms}},\underbrace{h\theta,0,\cdots,0}_{m \mathrm{\ terms}},h\theta^2,0,\cdots]^T$. We can see that $m$ only affects the position of the non-zero coefficients, which means that $m$ has no effect on calculating the gradients in Section \ref{sec:BFGS}. Moreover, in Section \ref{sec:szego}, the zero coefficients have no contribution to (\ref{eqn:tfunc}) which is the key component in the CRB analysis. Thus, assuming $m=1$ is without loss of generality.

The self-interference cancellation at the relay should be such that $|\theta|<1$ is possible with proper choice of $\alpha$. Such $\alpha$ keeps the system stable and guarantees finite average relay transmit power. The average relay transmit power is calculated as
\begin{align}\label{eqn:power_alpha}
{\rm E}[t_{\rm r}[n]t^*_{\rm r}[n]]&=\alpha^2\sum_{k=1}^{\infty}(\alpha^2|h_{\rm rr}|^2)^{(k-1)}\left(P_{\rm s}|h|^2+\sigma_{\rm r}^2\right)=\alpha^2\frac{P_{\rm s}|h|^2+\sigma_{\rm r}^2}{1-\alpha^2|h_{\rm rr}|^2}.
\end{align}
Define $P_{\rm r}$ as the maximum relay transmit power. The condition for the stability of the system is given by\cite{riihonen2009power}
\begin{align}\label{eqn:power_alpha1}
{\rm E}[t_{\rm r}[n]t^*_{\rm r}[n]]\leq P_{\rm r},
\end{align}
where the expectation in (\ref{eqn:power_alpha1}) is with respect to the noise. By solving (\ref{eqn:power_alpha1}), $\alpha$ should satisfy $\alpha^2|h_{\rm rr}|^2=|\theta|^2<1$. However, in a channel estimation scenario, the expectation value of $h_{\rm rr}$ is used instead of its instantaneous value in $\alpha$. We can choose $\alpha$ to satisfy a long term condition ${\rm E}[\alpha^2|h_{\rm rr}|^2]<1$ which leads to $\alpha^2\sigma_{\rm rr}^2<1$. The RSI variance $\sigma_{\rm rr}^2$ can be obtained at the pre-stage. \textcolor{blue}{Using the variance of the RSI channel instead of its realizations is a common problem in AF FD relays since the RSI is considered as the residual error which cannot be further estimated after all the self-interference cancellation approaches. Note that RSI channel realization might exceed some threshold. If that happens, $\alpha$ can be adjusted to make the relay transmit with its maximum power. This will lead to clipping and distortions but not instability. In addition, the clipping case happens with small probability since the RSI channel has small variance \cite{sabharwal2014band} which limits the dynamic range of the realizations. To further reduce the clipping probability, a fixed power margin between the relay gain power and the maximum power can be made on $\alpha$ to increase the threshold. Therefore, we do not incorporate these distortions in our system model and analysis.} We can choose $\alpha$ to first normalize the received signal power, and then to amplify the signal power to $P_{\rm r}$ which is the maximum power the relay can use given the power margin. Such an $\alpha$ is given by
\begin{align}\label{eqn:power_alpha2}
\alpha^2=\frac{P_{\rm r}}{P_{\rm s}+P_{\rm r}\sigma_{\rm rr}^2+\sigma_{\rm r}^2}\ \ .
\end{align}

Since the non-zero coefficients $\theta^{k-1}$ at the $k$th taps of the IIR channel decrease in amplitude with increasing tap index $k$, we can assume that most of the energy (e.g. 99\%) is contained in a finite length of the overall channel impulse response\cite{Kim2012full}. Define $L$ as the effective length of the overall impulse response which is $h[k]:=h\theta^{k},k=0,\cdots,L-1$. Thus, we use a block-based transmission with a guard time of $L$ symbol intervals to avoid inter-block interference\cite{li2016fulltwr}. At the receiver, it receives $N+L$ symbols and discards the last $L$ symbols. Without loss of generality, the block length $N$ is assumed to be far greater than $L$, so the rate loss due to the guard time is negligible. With the effective length $L$ and block-based transmission, we can truncate the IIR channel. 

Let $\boldsymbol{H}_{\theta}$ be the matrix form of the channel in one block, which is given by an $N \times N$ Toeplitz matrix with first column $[1,\theta,\theta^2,\cdots,\theta^{L-1},0,\cdots,0]^T$ and first row $[1,0,\cdots,0]$.

We rewrite the output in terms of $\boldsymbol{x}:=[x[0],\cdots,x[N-1]]^T$ and $\boldsymbol{y}:=[y_{\rm d}[1],\cdots,y_{\rm d}[N]]^T$ as:
\begin{align}\label{eqn:trainingdata}
\boldsymbol{y}=h\boldsymbol{H}_\theta\boldsymbol{x}+d\boldsymbol{H}_\theta\boldsymbol{n}_{\rm r}+\boldsymbol{n}_{\rm d},
\end{align}
where $\boldsymbol{n}_{\rm r}$ and $\boldsymbol{n}_{\rm d}$ are noise vectors composed of independent samples from the same distribution as $n_{\rm r}[n]$ and $n_{\rm d}[n]$ respectively. As can be seen from the matrix expression, $h\boldsymbol{H}_{\theta}$ is the overall channel and the sum of the last two terms in (\ref{eqn:trainingdata}) is the colored noise. Thus, the overall channel becomes an ISI channel. In (\ref{eqn:trainingdata}), we assume distortion of the signal caused by hardware impediment is negligible due to sufficient passive self-interference suppression and analog cancellation in RF\cite{Duarte12full}. However, if the distortion has to be considered, (\ref{eqn:trainingdata}) does not change because the distortion can be incorporated as part of the noise. To be specific, the distortion from the transmitter and the receiver are incorporated into the colored noise term $d\boldsymbol{H}_\theta\boldsymbol{n}_{\rm r}$ and additive noise term $\boldsymbol{n}_{\rm d}$ respectively\cite{Cirik16}. Since we have explicitly labeled the noise variances of $\boldsymbol{n}_{\rm r}$ and $\boldsymbol{n}_{\rm d}$ as $\sigma_{\rm r}^2$ and $\sigma_{\rm d}^2$ respectively, the incorporation of the distortion can be captured by modifying the noise variance values.

\section{Channel Estimation}\label{sec:channelest}
\subsection{Maximum Likelihood Formulation}
We now derive the ML estimator of $h$ and $\theta$ for a given training sequence $\boldsymbol{x}$. We are only interested in $h$ and $\theta$ since knowing them is enough for detection and equalization. In (\ref{eqn:trainingdata}) we have three parameters $h$, $\theta$ and $d$. The coefficients of the desired signal $\boldsymbol{x}$ is $h\boldsymbol{H}_\theta$ which only contain $h$ and $\theta$ while $d$ appears in the colored noise term $d\boldsymbol{H}_\theta \boldsymbol{n}_{\rm r}$. When detecting $\boldsymbol{x}$, $d$ is not necessary. For example, at high SNR, a zero-forcing detector can be used which is obtained by using $h$ and $\theta$ to calculate the inverse of $h\boldsymbol{H}_\theta$. We set $d$ as a nuisance parameter and integrate out the nuisance parameter from the likelihood function which is an established method\cite{MK1993} to deal with it in the likelihood function. We have
\begin{align}
p(\boldsymbol{y}|h,\theta)=\int p(\boldsymbol{y}|h,\theta,d)p(d)\mathrm{d}d.
\end{align}
Since $p(\boldsymbol{y}|h,\theta,d)$ and $p(d)$ are Gaussian distributed, it is shown in Appendix \rom{1} that the distribution of $p(\boldsymbol{y}|h,\theta)$ is also Gaussian with mean and covariance matrix 
\begin{align}
\boldsymbol{\mu}=h\boldsymbol{H}_\theta,\boldsymbol{C}=\alpha^2\sigma_{\rm r}^2\boldsymbol{H}_\theta\boldsymbol{H}_\theta^H+\sigma_{\rm d}^2\boldsymbol{I}_N.
\end{align}
Therefore, the likelihood function of $\boldsymbol{y}$ is
\begin{align}
&p(\boldsymbol{y}|h,\theta)=\frac{1}{\pi^N|\boldsymbol{C}|}\exp\left(-(\boldsymbol{y}-\boldsymbol{\mu})^H\boldsymbol{C}^{-1}(\boldsymbol{y}-\boldsymbol{\mu})\right),
\end{align}
where $|\boldsymbol{C}|$ denotes the determinant of matrix $\boldsymbol{C}$. The corresponding log-likelihood function is\cite{MK1993}
\begin{align}\label{eqn:loglikelihood}
\log p(\boldsymbol{y}|h,\theta)=&-N\log\pi-\log|\boldsymbol{C}|-(\boldsymbol{y}-\boldsymbol{\mu})^H\boldsymbol{C}^{-1}(\boldsymbol{y}-\boldsymbol{\mu}).
\end{align}

Maximizing (\ref{eqn:loglikelihood}) is equivalent to minimizing the last two terms of it. Let $f$ denote our objective function. 
\begin{align}\label{eqn:obf}
f(h,\theta)=\log|\boldsymbol{C}|+(\boldsymbol{y}-\boldsymbol{\mu})^H\boldsymbol{C}^{-1}(\boldsymbol{y}-\boldsymbol{\mu}).
\end{align}
The ML estimator is given by
\begin{align}\label{eqn:mlestimator}
\{\hat{h},\hat{\theta}\}=\argmin_{h,\theta}&\big\{\log|\boldsymbol{C}|+(\boldsymbol{y}-\boldsymbol{\mu})^H\boldsymbol{C}^{-1}(\boldsymbol{y}-\boldsymbol{\mu})\big\}.
\end{align}

Note that the two parameters are complex. We denote $h=h_x+jh_y$ where $h_x$ and $h_y$ are the real part and imaginary part of $h$ respectively and $j$ is the imaginary unit. Similarly we have $\theta=\theta_x+j\theta_y$. Before we solve the ML estimator, we will simplify the objective function to express it in terms of only one complex parameter $\theta$. First we take derivative of $f$ with respect to $h$,
\begin{align}
\frac{\partial f}{\partial h}=-\boldsymbol{y}^H\boldsymbol{C}^{-1}\boldsymbol{H}_{\theta}\boldsymbol{x}+h^*\boldsymbol{x}^H\boldsymbol{H}_{\theta}^H\boldsymbol{C}^{-1}\boldsymbol{H}_{\theta}\boldsymbol{x}.
\end{align}
Setting the derivative to 0 we have
\begin{align}\label{eqn:h}
h=(\boldsymbol{x}^H\boldsymbol{H}_{\theta}^H\boldsymbol{C}^{-1}\boldsymbol{H}_{\theta}\boldsymbol{x})^{-1}\boldsymbol{x}^H\boldsymbol{H}_{\theta}^H\boldsymbol{C}^{-1}\boldsymbol{y}.
\end{align}
We can substitute (\ref{eqn:h}) into (\ref{eqn:obf}) to eliminate $h$. 

The objective function is not convex with respect to $\theta$. To solve the problem numerically, we use the Broyden-Fletcher-Goldfarb-Shanno (BFGS) algorithm\cite{Kevin2012learning}, which is a popular quasi-Newton method. Note that the constraint $|\theta|<1$ is imposed to ensure stability in (\ref{eqn:rec_sig1}). Euclidean projection is further applied to $\hat{\theta}$ to ensure that $|\hat{\theta}|<1$ so that the estimates conform with the stability assumption. Because the algorithm can only deal with real valued parameters, the real and imaginary parts are optimized separately.

\subsection{BFGS algorithm}\label{sec:BFGS}
We use the BFGS algorithm which is also used in \cite{li2016fulltwr} to solve the ML estimator in a different two-way relay context. The algorithm needs the gradients of $f$ with respect to $\theta_x$ and $\theta_y$. We derive the gradients in Appendix \rom{3}. A linear MMSE estimator is used to initialize the BFGS algorithm, which helps the algorithm to converge faster and to reduce the possibility of trapping in a local minimum. We now elaborate on the initialization before we provide the details of the BFGS algorithm.

Pairs of received samples can be used for linear MMSE estimation even though the received samples $\boldsymbol{y}$ are not linear in the desired parameters $h$ and $\theta$. We take two received symbols $y_{\rm d}[2]$ and $y_{\rm d}[3]$ to estimate $h$ and $\theta$. For estimating $h$, the second received symbol at the destination is used, which is
\begin{eqnarray}
y_{\rm d}[2]=hx[1]+dn_{\rm r}[1]+n_{\rm d}[2].
\end{eqnarray}
The linear MMSE estimator $\hat{h}_0$ is given by [30, Sec. 12.3]
\begin{eqnarray}\label{eqn:mmse1}
\hat{h}_0=\frac{\alpha^2\sigma^2_{\rm sr}\sigma^2_{\rm rd}x^*[1]y_{\rm d}[2]}{\alpha^2\sigma^2_{\rm sr}\sigma^2_{\rm rd}|x[1]|^2+\alpha^2\sigma^2_{\rm rd}\sigma_{\rm r}^2+\sigma_{\rm d}^2}.
\end{eqnarray}
Let $\tilde{h}_0$ be the residual estimation error of $h$, i.e. $h=\hat{h}_0+\tilde{h}_0$. Thus, $\tilde{h}_0$ is a random variable with zero mean and variance $\sigma^2_{\tilde{h}_0}$ which is given by
\begin{eqnarray}
\sigma^2_{\tilde{h}_0}=\frac{\alpha^2\sigma^2_{\rm sr}\sigma^2_{\rm rd}(\alpha^2\sigma^2_{\rm rd}\sigma_{\rm r}^2+\sigma_{\rm d}^2)}{\alpha^2\sigma^2_{\rm sr}\sigma^2_{\rm rd}|x[1]|^2+\alpha^2\sigma^2_{\rm rd}\sigma_{\rm r}^2+\sigma_{\rm d}^2}.
\end{eqnarray}

After having $\hat{h}_0$, we can estimate $\theta$. First we use $\hat{h}_0$ to remove the known part $\hat{h}_0x[2]$ in $y_{\rm d}[3]$. The remaining signal of $y_{\rm d}[3]$ is as follows:
\begin{align}
y'_{\rm d}[3]=&\hat{h}_0\theta x[1]+\tilde{h}_0\theta x[1]+\tilde{h}_0x[2]+hdn_{\rm r}[1]+dn_{\rm r}[2]+n_{\rm d}[3].
\end{align}
The linear MMSE estimator of $\theta$ is
\begin{eqnarray}\label{eqn:mmse2}
\hat{\theta}_0=\frac{\hat{h}_0\alpha^2\sigma^2_{\rm rr}x^*[1]y'_{\rm d}[3]}{\hat{h}_0^2\alpha^2\sigma^2_{\rm rr}|x[1]|^2+\sigma^2_{\tilde{h}_0}|x[2]|^2+\sigma^2_{\tilde{h}_0}\alpha^2\sigma_{\rm rr}^2|x[1]|^2+\alpha^4\sigma_{\rm sr}^2\sigma_{\rm rd}^4\sigma_{\rm r}^2+\alpha^2\sigma_{\rm rd}^2\sigma_{\rm r}^2+\sigma_{\rm d}^2}.
\end{eqnarray}

Though we only make use of one training symbol in the above linear MMSE method, it is possible to extend the method to use multiple symbols, in which case a special training sequence with $L-1$ zeros followed by one symbol is transmitted, where $L$ is the effective length of the channel impulse response. 

We now provide the BFGS algorithm which uses the initialization explained above, and the gradients in Appendix \rom{3}.
\vspace{-5mm}
\begin{center}
\line(1,0){450} 
\end{center}
\vspace{-5mm}
\begin{tabbing}
Initialize:\ \= $\boldsymbol{z}_{0}\triangleq[\hat{\theta}_x\ \hat{\theta}_y]^T$,  \ $\boldsymbol{A}_{0}^{-1}=\boldsymbol{I}_{2\times2}$.\\
Repeat until convergence for $k$: (BFGS)\\
\> 1. Obtain a search direction $\boldsymbol{p}_{k}=-\boldsymbol{A}_{k}^{-1}\nabla f(\boldsymbol{z}_{k})$.\\
\> 2. Find stepsize $\lambda_k$ by backtracking linesearch, then update $\boldsymbol{z}_{k+1}=\boldsymbol{z}_{k}+\lambda_k\boldsymbol{p}_{k}$.\\
\> 3. Set $\boldsymbol{s}_{k}=\lambda_k\boldsymbol{p}_{k}$, $\boldsymbol{v}_{k}=\nabla f(\boldsymbol{z}_{k+1})-\nabla f(\boldsymbol{z}_{k})$\\
\> 4. Update the inverse Hessian approximation by\\
\> $\boldsymbol{A}_{k+1}^{-1}=\boldsymbol{A}_{k}^{-1}+\frac{(\boldsymbol{s}_{k}^T\boldsymbol{v}_{k}+\boldsymbol{v}_{k}^T\boldsymbol{A}_{k}^{-1}\boldsymbol{v}_{k})\boldsymbol{s}_{k}\boldsymbol{s}_{k}^T}{(\boldsymbol{s}_{k}^T\boldsymbol{v}_{k})^2}-\frac{\boldsymbol{A}_{k}^{-1}\boldsymbol{v}_{k}\boldsymbol{s}_{k}^T+\boldsymbol{s}_{k}\boldsymbol{v}_{k}^T\boldsymbol{A}_{k}^{-1}}{\boldsymbol{s}_{k}^T\boldsymbol{v}_{k}}$\\
Obtain the converged result $\boldsymbol{z}_{k}$ and construct the estimate $\hat{\theta}$ from it.\\
If $|\hat{\theta}|>1$, $\hat{\theta}=\hat{\theta}/|\hat{\theta}|$ (Euclidean projection).
\end{tabbing}
\vspace{-10mm}
\begin{center}
\line(1,0){450}
\end{center}

After the initial values are input, the complex parameter $\theta$ is optimized by the BFGS algorithm. The iteration is controlled by the index $k$. The results of one iteration will be used as initial values for the next iteration. Let $\hat{\theta}$ be the estimate of $\theta$ and is obtained from $\boldsymbol{z}_{k}$. In particular, there is a constraint $|\theta|<1$ on $\theta$. We use Euclidean projection, which in this case is a vector normalization, to keep $\hat{\theta}$ in its valid region. If the result of $\hat{\theta}$ is a point outside of the valid region, Euclidean projection maps the outside point to its nearest valid point. The BFGS algorithm is able to converge since it uses the Hessian approximation matrix to update the search direction. The positive definite property of the Hessian approximation matrix implies a descent search direction, which guarantees convergence [31, Sec. 8.3.5]. However, due to the non-convexity of the objective function, the algorithm might be trapped in a local minimum. To avoid this, we use MMSE estimates of the parameters to initialize the algorithm as mentioned above.

The complexity of the algorithm is dominated by the matrix inversion of the covariance matrix $\boldsymbol{C}$ in the calculation of the gradients and the objective function (\ref{eqn:obf}). For large training length $N$, $\boldsymbol{C}$ asymptotically becomes to a positive definite Toeplitz matrix. The complexity of inverting it is $O(N\log^2 N)$\cite{MK1993}. The approximate inverse-Hessian matrix update only depends on the number of parameters to be estimate but not on $N$. Therefore, the total complexity of the BFGS algorithm in one iteration is $O(N\log^2 N)$ for large $N$. Moreover, the algorithm with linear MMSE initialization converges faster than random initialization based on our observation in the simulation. Thus, our initialization method also helps to reduce the complexity of the algorithm.

\section{Optimal Training Sequences}\label{sec:CRB}
\subsection{Cramer-Rao Bounds}
The CRB is derived not only to show the accuracy of the estimates but also to act as a metric when designing the training sequences. Differentiating the log-likelihood function $\log p(\boldsymbol{y}|h,\theta)$ twice, we can obtain the Fisher information matrix (FIM). Let $\boldsymbol{\xi}=[h\ \theta]^T$ be the vector of parameters. The FIM is given by
\begin{align}
\boldsymbol{\Gamma}(\boldsymbol{\xi})=\mathrm{E}\left[\frac{\partial \log p}{\partial \boldsymbol{\xi}^*}\frac{\partial \log p}{\partial \boldsymbol{\xi}^T}\right].
\end{align}
The $(m,n)$ element of $\boldsymbol{\Gamma}$ is given by 
\begin{align}
\Gamma_{mn}=\frac{\partial\boldsymbol{\mu}^H}{\partial \xi_m^*}\boldsymbol{C}^{-1}\frac{\partial\boldsymbol{\mu}}{\partial \xi_n}+{\rm tr}\left(\boldsymbol{C}^{-1}\frac{\partial\boldsymbol{C}}{\partial \xi_m^*}\boldsymbol{C}^{-1}\frac{\partial\boldsymbol{C}}{\partial \xi_n}\right),
\end{align}
where $\xi_m$ is the $m$th element of $\boldsymbol{\xi}$. Thus we have
\begin{align}
\Gamma_{11}&=\frac{\partial\boldsymbol{\mu}^H}{\partial h^*}\boldsymbol{C}^{-1}\frac{\partial\boldsymbol{\mu}}{\partial h}+{\rm tr}\left(\boldsymbol{C}^{-1}\frac{\partial\boldsymbol{C}}{\partial h^*}\boldsymbol{C}^{-1}\frac{\partial\boldsymbol{C}}{\partial h}\right)=\boldsymbol{x}^H\boldsymbol{H}_{\theta}^H\boldsymbol{C}^{-1}\boldsymbol{H}_{\theta}\boldsymbol{x},\\
\Gamma_{22}&=\frac{\partial\boldsymbol{\mu}^H}{\partial \theta^*}\boldsymbol{C}^{-1}\frac{\partial\boldsymbol{\mu}}{\partial \theta}+{\rm tr}\left(\boldsymbol{C}^{-1}\frac{\partial\boldsymbol{C}}{\partial \theta^*}\boldsymbol{C}^{-1}\frac{\partial\boldsymbol{C}}{\partial \theta}\right)\nonumber\\
&=|h|^2\boldsymbol{x}^H\boldsymbol{B}_{\theta}^H\boldsymbol{C}^{-1}\boldsymbol{B}_{\theta}\boldsymbol{x}+\alpha^4\sigma_{\rm r}^4{\rm tr}\left(\boldsymbol{C}^{-1}\boldsymbol{H}_{\theta}\boldsymbol{B}_{\theta}^H\boldsymbol{C}^{-1}\boldsymbol{B}_{\theta}\boldsymbol{H}_{\theta}^H\right).\label{eqn:crbtheta}
\end{align}
Since $\boldsymbol{C}$ is not a function of $h$,
\begin{align}
\Gamma_{12}=\frac{\partial\boldsymbol{\mu}^H}{\partial h^*}\boldsymbol{C}^{-1}\frac{\partial\boldsymbol{\mu}}{\partial \theta}
=h\boldsymbol{x}^H\boldsymbol{H}_\theta^H\boldsymbol{C}^{-1}\boldsymbol{B}_\theta\boldsymbol{x},\ \ \Gamma_{21}=\frac{\partial\boldsymbol{\mu}^H}{\partial \theta^*}\boldsymbol{C}^{-1}\frac{\partial\boldsymbol{\mu}}{\partial h}=h^*\boldsymbol{x}^H\boldsymbol{B}_\theta^H\boldsymbol{C}^{-1}\boldsymbol{H}_\theta\boldsymbol{x}.
\end{align}

The CRB is given by the trace of the inverse of $\boldsymbol{\Gamma}$, which is $CRB_{\boldsymbol{\xi}}=\mathrm{tr}(\boldsymbol{\Gamma}^{-1})$. In particular, the CRBs for each parameter are the diagonal elements of the inverse FIM. Since  $\boldsymbol{\Gamma}$ is a 2 by 2 complex matrix, we can find its inverse by calculating its determinant and adjoint. The determinant is $|\boldsymbol{\Gamma}|=\Gamma_{11}\Gamma_{22}-\Gamma_{12}\Gamma_{21}$. Therefore, the CRBs are given by 
\vspace{-5mm}
\begin{align}\label{eqn:crbexact}
CRB_h=\Gamma_{22}/|\boldsymbol{\Gamma}|, \ \ \ CRB_\theta=\Gamma_{11}/|\boldsymbol{\Gamma}|
\end{align}
\vspace{-15mm}
\subsection{Training Sequence Design via the CRB}\label{sec:szego}
In this subsection, we analyze the CRB by using theorems for inverses, products, and eigenvalues of Toeplitz matrices derived in \cite{Gray2006} based on Szeg\"{o}'s theorem about asymptotic behavior of Toeplitz matrix eigenvalues. The CRB is minimized in the regime where the training length $N$ is large. We show that the optimal training sequence that minimizes the CRB is sinusoidal and we characterize the frequency of this sinusoidal. The key idea behind this is that circulant matrices have sinusoidal eigenvectors and the Toeplitz matrices in the CRB expression can be well approximated by circulant matrices for large $N$.

To analyze the asymptotic behavior of Toeplitz matrices, we define an $N\times N$ Toeplitz matrix $\boldsymbol{T}_N$ whose elements $t_k$ satisfy $\sum_{k=-\infty}^{\infty}|t_k|<\infty$. According to \cite{Gray2006}, $\boldsymbol{T}_N$ is equivalent to a circulant matrix as $N\rightarrow\infty$, and can be expressed as $\boldsymbol{T}_N(t(\lambda))$ where $t(\lambda)=\sum_{k=-\infty}^{\infty}t_ke^{j\lambda k}$. Now we show that $\boldsymbol{H}_\theta$ is asymptotically equivalent to $\boldsymbol{T}_N(t(\lambda))$. First, since both $\boldsymbol{H}_\theta$ and $\boldsymbol{T}_N(t(\lambda))$ are banded Toeplitz matrices [32, Sec. 4.3], their strong norms (operator norms) are bounded. Secondly, let $t_k=\theta^k$ for $k=0,\cdots, L-1$ and otherwise $t_k=0$, we have $\lim_{N \to \infty}||\boldsymbol{H}_\theta-\boldsymbol{T}_N(t(\lambda))||=0$, where $||\boldsymbol{A}||$ denotes the weak norm (Hilbert-Schmidt norm) of matrix $\boldsymbol{A}$. With the two conditions above, we can say that $\boldsymbol{H}_\theta$ and $\boldsymbol{T}_N(t(\lambda))$ are asymptotically equivalent [32, Sec. 2.3].  Therefore, we will write $\boldsymbol{H}_\theta=\boldsymbol{T}_N(t(\lambda))$ which will be understood to hold for asymptotically large $N$ and thus the asymptotic properties which are introduced later can be applied to analyze the CRB. We have the following expression for $t(\lambda)$,
\begin{align}\label{eqn:tfunc}
t(\lambda)=\sum_{k=0}^{L-1}\theta^k e^{j\lambda k}=\frac{1-|\theta|^L}{1-\theta e^{j\lambda}}=\frac{1}{1-\theta e^{j\lambda}},
\end{align}
where the assumption of channel energy $|\theta|^L \approx 0$ is used. The covariance matrix $\boldsymbol{C}$ is 
\begin{align}
\boldsymbol{C}=\alpha^2\sigma_{\rm r}^2\boldsymbol{T}_N(t(\lambda))\boldsymbol{T}_N(t^*(\lambda))+\sigma_{\rm d}^2\boldsymbol{I}_N.
\end{align}
Without loss of generality, we set $\sigma_{\rm r}^2=\sigma_{\rm d}^2=1$. According to \cite{Gray2006}, the product of two Toeplitz matrices is a Toeplitz matrix asymptotically, as well as the inverse of a Toeplitz matrix. Thus we have
\begin{align}
\boldsymbol{C}\approx\alpha^2\boldsymbol{T}_N(|t(\lambda)|^2)+\boldsymbol{I}_N=\boldsymbol{T}_N(\alpha^2|t(\lambda)|^2+1),\ \ \ \boldsymbol{C}^{-1}\approx\boldsymbol{T}_N\left(\frac{1}{\alpha^2|t(\lambda)|^2+1}\right).
\end{align}

The Fisher information of the source-relay-destination channel $h$ is 
\begin{align}
\Gamma_{11}&=\boldsymbol{x}^H\boldsymbol{H}_{\theta}^H\boldsymbol{C}^{-1}\boldsymbol{H}_{\theta}\boldsymbol{x}=\boldsymbol{x}^H\boldsymbol{T}_N(t^*(\lambda))\boldsymbol{T}_N\left(\frac{1}{\alpha^2|t(\lambda)|^2+1}\right)\boldsymbol{T}_N(t(\lambda))\boldsymbol{x}\\
&\approx\boldsymbol{x}^H\boldsymbol{T}_N\left(\frac{|t(\lambda)|^2}{\alpha^2|t(\lambda)|^2+1}\right)\boldsymbol{x}=\frac{|t(\lambda)|^2}{\alpha^2|t(\lambda)|^2+1}||\boldsymbol{x}||^2,\label{eqn:F11T}
\end{align}
where $\frac{|t(\lambda)|^2}{\alpha^2|t(\lambda)|^2+1}$ is the eigenvalue of $\boldsymbol{T}_N\left(\frac{|t(\lambda)|^2}{\alpha^2|t(\lambda)|^2+1}\right)$ and depends on $\lambda$. Similarly, we can denote $\boldsymbol{B}_\theta=\boldsymbol{T}_N(g(\lambda))$ where $g(\lambda)=\frac{e^{j\lambda}}{(1-\theta e^{j\lambda})^2}$ is the derivative of $t(\lambda)$ with respect to $\theta$. The Fisher information of the RSI channel can be represented by Toeplitz matrices as
\begin{align}
\Gamma_{22}&=|h|^2\boldsymbol{x}^H\boldsymbol{B}_{\theta}^H\boldsymbol{C}^{-1}\boldsymbol{B}_{\theta}\boldsymbol{x}+\alpha^4{\rm tr}\left(\boldsymbol{C}^{-1}\boldsymbol{H}_{\theta}\boldsymbol{B}_{\theta}^H\boldsymbol{C}^{-1}\boldsymbol{B}_{\theta}\boldsymbol{H}_{\theta}^H\right)\\
&\approx|h|^2\boldsymbol{x}^H\boldsymbol{T}_N\left(\frac{|g(\lambda)|^2}{\alpha^2|t(\lambda)|^2+1}\right)\boldsymbol{x}+\alpha^4{\rm tr}\left(\boldsymbol{T}_N\left(\frac{|g(\lambda)|^2|t(\lambda)|^2}{(\alpha^2|t(\lambda)|^2+1)^2}\right)\right)\\
&=|h|^2\frac{|g(\lambda)|^2}{\alpha^2|t(\lambda)|^2+1}||\boldsymbol{x}||^2+\alpha^4\frac{||\boldsymbol{x}||^2}{2\pi P_{\rm s}}\int_{0}^{2\pi}\frac{|g(\lambda)|^2|t(\lambda)|^2}{(\alpha^2|t(\lambda)|^2+1)^2}{\rm d}\lambda .\label{eqn:Gamma22}
\end{align}
We can simplify the first term in (\ref{eqn:Gamma22}) similarly to $\Gamma_{11}$. The second term comes from the fact that the trace of Toeplitz matrices is equal to the integral of the function of $\lambda$ that characterizes it\cite{Gray2006}. Simplifying this integral we have
\begin{align}\label{eqn:int1}
\int_{0}^{2\pi}\frac{|g(\lambda)|^2|t(\lambda)|^2}{(\alpha^2|t(\lambda)|^2+1)^2}{\rm d}\lambda=\int_{0}^{2\pi}\frac{1}{|1-\theta e^{j\lambda}|^2(\alpha^2+|1-\theta e^{j\lambda}|^2)^2}{\rm d}\lambda.
\end{align}
In our FD relay system, we assume $P_{\rm r} \gg P_{\rm s}$ since $P_{\rm s}$ is the transmit power at the source which incorporates the path loss. Thus $\alpha^2 \gg 1$. Note that since $|\theta|<1$, we have $\alpha^2 \gg |1-\theta e^{j\lambda}|^2$. We can approximate the integral as
\begin{align}
\frac{1}{2\pi}\int_{0}^{2\pi}\frac{1}{|1-\theta e^{j\lambda}|^2(\alpha^2+|1-\theta e^{j\lambda}|^2)^2}{\rm d}\lambda\approx \frac{1}{2\pi}\int_{0}^{2\pi}\frac{1}{|1-\theta e^{j\lambda}|^2\alpha^4}{\rm d}\lambda=\frac{1}{\alpha^4}\frac{1}{(|\theta|+1)\big||\theta|-1\big|}.\label{eqn:int2}
\end{align}
Therefore, the Fisher information of the RSI channel $\theta$ becomes
\begin{align}\label{eqn:F33T}
\Gamma_{22}&=\frac{|h|^2|g(\lambda)|^2}{\alpha^2|t(\lambda)|^2+1}||\boldsymbol{x}||^2+\frac{1}{P_{\rm s}(|\theta|+1)\big||\theta|-1\big|}||\boldsymbol{x}||^2.
\end{align}
Similarly, we can represent $\Gamma_{12}$ and $\Gamma_{21}$ as
\begin{align}
\Gamma_{12}&=h\boldsymbol{x}^H\boldsymbol{B}_{\theta}^H\boldsymbol{C}^{-1}\boldsymbol{H}_{\theta}\boldsymbol{x}\approx\boldsymbol{x}^H\boldsymbol{T}_N\left(\frac{ht(\lambda)g^*(\lambda)}{\alpha^2t(\lambda)|^2+1}\right)\boldsymbol{x},\\
\Gamma_{21}&=h^*\boldsymbol{x}^H\boldsymbol{H}_{\theta}^H\boldsymbol{C}^{-1}\boldsymbol{B}_{\theta}\boldsymbol{x}\approx\boldsymbol{x}^H\boldsymbol{T}_N\left(\frac{h^*t^*(\lambda)g(\lambda)}{\alpha^2|t(\lambda)|^2+1}）\right)\boldsymbol{x}.
\end{align}
To calculate the CRB, we need the product of $\Gamma_{12}$ and $\Gamma_{21}$ which is
\begin{align}\label{eqn:prod1221}
\Gamma_{12}\Gamma_{21}=|h|^2\frac{p^2(\lambda)}{(\alpha^2|t(\lambda)|^2+1)^2}||\boldsymbol{x}||^4.
\end{align}
where $p(\lambda)=\frac{1}{2}[t^*(\lambda)g(\lambda)+t(\lambda)g^*(\lambda)]$. Function $p(\lambda)$ is the real part of $t^*(\lambda)g(\lambda)$ and it shows that only the symmetric part of the Toeplitz matrix affects the product. Thus, the CRB of $\theta$ is 
\begin{align}\label{eqn:crb_theta}
CRB_\theta=\frac{\Gamma_{11}}{\Gamma_{11}\Gamma_{22}-\Gamma_{12}\Gamma_{21}}=\frac{1}{\Gamma_{11}\Gamma_{22}-\Gamma_{12}\Gamma_{21}}\boldsymbol{x}^H\boldsymbol{T}_N(\frac{|t(\lambda)|^2}{\alpha^2|t(\lambda)|^2+1})\boldsymbol{x}.
\end{align}

Minimizing (\ref{eqn:crb_theta}) is to find a eigenvalue of $\boldsymbol{T}_N(\frac{|t(\lambda)|^2}{\alpha^2|t(\lambda)|^2+1})$ which depends on $\lambda$. Note that the term $\Gamma_{11}\Gamma_{22}-\Gamma_{12}\Gamma_{21}$ also depends on $\lambda$, the minimization of (\ref{eqn:crb_theta}) is only through $\lambda$ and the optimal training sequence is the corresponding eigenvector. Since $\boldsymbol{T}_N(\frac{|t(\lambda)|^2}{\alpha^2|t(\lambda)|^2+1})$ is asymptotically equivalent to a circulant matrix, the eigenvector is sinusoidal. Plug (\ref{eqn:F11T}), (\ref{eqn:F33T}), and (\ref{eqn:prod1221}) into (\ref{eqn:crb_theta}), 
\begin{align}
CRB_\theta=\frac{1}{||\boldsymbol{x}||^2}\frac{|t(\lambda)|^2(\alpha^2|t(\lambda)|^2+1)}{|h|^2|t(\lambda)|^2|g(\lambda)|^2+A|t(\lambda)|^2(\alpha^2|t(\lambda)|^2+1)-|h|^2|p(\lambda)|^2}\triangleq\frac{1}{||\boldsymbol{x}||^2}F(\lambda),\label{eqn:crb_asymp1}
\end{align}
where $A=(P_{\rm s}(|\theta|+1)\big||\theta|-1\big|)^{-1}$. To find the frequency of the sinusoidal training sequence, we minimize the $F(\lambda)$ in (\ref{eqn:crb_asymp1}) with $\lambda\in[0,2\pi]$. Simplifying $F(\lambda)$ we have
\begin{align}
F(\lambda)=\frac{\alpha^2+|1-\theta e^{j\lambda}|^2}{\frac{|h|^2}{|1-\theta e^{j\lambda}|^2}+A(\alpha^2+|1-\theta e^{j\lambda}|^2)-\frac{1}{4}|h|^2\frac{({\rm Re}[e^{j\lambda}-\theta^*])^2}{|1-\theta e^{j\lambda}|^4}}.
\end{align}	
Let $z=|1-\theta e^{j\lambda}|^2$ and $z\in[(1-|\theta|^2),(1+|\theta|)^2]$. Substitute it into $F(\lambda)$, we have
\begin{align}\label{eqn:gx}
G(z)=\frac{\alpha^2+z}{\frac{|h|^2}{z}+A(\alpha^2+z)-\frac{|h|^2}{4z^2}\left(\frac{(1+|\theta|^2-z)\theta_{x}}{2|\theta|^2}+\sqrt{1-\frac{(1+|\theta|^2-x)^2}{4|\theta|^2}}\frac{\theta_{y}}{|\theta|}-\theta_{x}\right)^2},
\end{align}
where $\theta_{x}$ and $\theta_{y}$ are the real and imaginary parts of $\theta$ respectively.

The optimal solution that minimizes $F(\lambda)$ can be found by numerically solving $G'(z)=0$ which can be rewritten as a polynomial in $z$ with highest order 8. The coefficients of the polynomial are given in Appendix \rom{2}. Note that $z\in[(1-|\theta|^2),(1+\theta)^2]$, so that the two endpoints of the interval are also candidates for the optimal solution in case that the only solution to $G'(z)=0$ is a saddle point or there is no solution in the interval. After we get all the candidates (of which there are a maximum of 8), we are able to substitute each of them into $G(z)$ to find the one that minimizes the function. Since the Toeplitz matrices asymptotically behave equivalently to circulant matrices according to Lemma 4.2 in \cite{Gray2006}, the same way for circulant matrices can be used to find the corresponding eigenvector for Toeplitz matrices. Assume the solution that minimizes $CRB_\theta$ is $\lambda^*$, the corresponding optimal training sequence is given by $\frac{1}{\sqrt{N}}[1,\cdots,e^{j2\pi k\lambda^*},\cdots,e^{j2\pi(N-1)\lambda^*}]^T$ for $k=0,\cdots,N-1$. 

The CRB for $h$ can be also derived the same way as the CRB of $\theta$,
\begin{align}\label{eqn:crbh}
CRB_h=\frac{1}{||\boldsymbol{x}||^2}\frac{|h|^2|g(\lambda)|^2(\alpha^2|t(\lambda)|^2+1)+A(\alpha^2|t(\lambda)|^2+1)^2}{|h|^2|t(\lambda)|^2|g(\lambda)|^2+A|t(\lambda)|^2(\alpha^2|t(\lambda)|^2+1)-|h|^2|p(\lambda)|^2}.
\end{align}
It can also be minimized by finding the roots of a polynomial. Note that when minimizing the CRB for $\theta$, it is not guaranteed that the CRB of $h$ is minimized as well. However, we can minimize the sum of CRBs of $\theta$ and $h$ if both parameters are considered, also through polynomial rooting.  

The optimal training sequence depends on both of the channel $h$ and $\theta$ through $\lambda$. In practice, we do not have the information of $h$ and $\theta$ until the first training sequence is sent. We can apply an adaptive training method where the optimal training sequence is designed by using estimates obtained from its previous training sequence. In what follows we show through an approximation that the minimizer of (\ref{eqn:gx}) only weakly depends on $h$.  

\subsection{Low Complexity Approximation}
The optimal solution can be found by minimizing the CRB numerically via finding the polynomial roots. However, the complexity can be reduced by a certain approximation which we now describe. This provides an approximately optimal and practical solution for the problem. Assume $|\theta|$ is small, so that the value of $x$ is very close to 1. Then we can have the following approximation
\begin{align}
\left(\frac{(1+|\theta|^2-z)\theta_{x}}{2|\theta|^2}+\sqrt{1-\frac{(1+|\theta|^2-z)^2}{4|\theta|^2}}\frac{\theta_{y}}{|\theta|}-\theta_{x}\right)^2\approx 1 \ .
\end{align}
Thus,
\begin{align}\label{eqn:gx2}
G(z)\approx\frac{\alpha^2+z}{\frac{|h|^2}{z}+A(\alpha^2+z)-\frac{|h^2|}{4z^2}}\ .
\end{align}
Solving $G'(x)=0$ is equivalent to solving the following equation:
\begin{align}\label{eqn:gx3}
8|h^2|z^3+(4\alpha^2-3)|h|^2z^2-2\alpha^2|h|^2z=0.
\end{align}
Equation (\ref{eqn:gx3}) shows that $h$ does not affect the solution of $G'(z)=0$. One can verify that none of the three real roots of (\ref{eqn:gx3}) is in the valid interval of $z$ which is $[(1-|\theta|)^2,(1+|\theta|)^2]$. Note that $G(z)$ is an increasing function since $|h|^2>0$. Therefore the approximately optimal solution is the left endpoint of the interval i.e. $z=(1-|\theta|)^2$. Thus $\lambda=-\angle\theta$ where $\angle$ represents the phase of a complex number. Moreover, the channel $|h|$ does not affect the solution of $z$, so that the training sequence for estimating $\theta$ only depends on $\theta$ and not $|h|$, making it easier to implement than the optimal training sequence. The normalized corresponding training sequence is given by $\frac{1}{\sqrt{N}}[1,\cdots,e^{j2\pi k\lambda_1^*},\cdots,e^{j2\pi(N-1)\lambda_1^*}]^T$ for $k=0,\cdots,N-1$ where $\lambda_1^*$ minimizes (\ref{eqn:gx2}).
\section{Frequency-selective channels}\label{sec:freq}
In this section, we extend our channel estimation method to the case where the channels between nodes are frequency-selective fading. We show that our training method and CRB calculation can be extended to this case based on our analysis of the basic one-way relay system.

We assume the source-to-relay and relay-to-destination channels are frequency-selective fading with channel taps $\boldsymbol{h}_{\rm sr}=[h_{\rm sr}[1],h_{\rm sr}[2],\cdots,h_{\rm sr}[L_1]]$ and $\boldsymbol{h}_{\rm rd}=[h_{\rm rd}[1],h_{\rm rd}[2],\cdots,h_{\rm rd}[L_2]]$ respectively. Therefore, with block based transmission, the channel matrix for the source-to-relay channel $\boldsymbol{H}_{\rm sr}$ is an $N\times N$ Toeplitz matrix with first column $[\boldsymbol{h}_{\rm sr}^T,0,\cdots,0]^T$ and first row $[1,0,0,\cdots,0]$. For the relay-to-destination channel, the channel matrix $\boldsymbol{H}_{\rm rd}$ is also an $N\times N$ Toeplitz matrix with first column $[\boldsymbol{h}_{\rm rd}^T,0,\cdots,0]^T$ and first row $[1,0,0,\cdots,0]$. The received signal for the training phase is similar to (\ref{eqn:trainingdata}) and becomes
\begin{align}\label{eqn:training1}
\boldsymbol{y}_{\rm f}&=\alpha_{\rm f}\boldsymbol{H}_{\rm rd}\boldsymbol{H}_{\theta}\boldsymbol{H}_{\rm sr}\boldsymbol{x}+\alpha_{\rm f}\boldsymbol{H}_{\rm rd}\boldsymbol{H}_\theta\boldsymbol{n}_{\rm r}+\boldsymbol{n}_{\rm d},
\end{align}
where $\alpha_{\rm f}$ is the new power scaling factor for the frequency-selective channel given by
\begin{align}\label{eqn:power_alpha3}
\alpha_{\rm f}^2=\frac{P_{\rm r}}{P_{\rm s}(\sum_{i=1}^{L_1}\sigma_{{\rm sr}i}^2)+P_{\rm r}\sigma_{\rm rr}^2+\sigma_{\rm r}^2}\ \ .
\end{align}
where $\sigma_{{\rm sr}i}^2$ is the variance of the $i$th source-to-relay channel tap. The overall channel is $\alpha_{\rm f}\boldsymbol{H}_{\rm rd}\boldsymbol{H}_{\theta}\boldsymbol{H}_{\rm sr}$ in the frequency-selective case instead of $h\boldsymbol{H}_{\theta}$ for flat fading. 

To extend our training method, we can use the theorem for products of Toeplitz matrices which is explained in Section \ref{sec:szego} to approximate the overall channel matrix as a Toeplitz matrix. According to the theorem, when the training length is large, the product of two Toeplitz matrices is still a Toeplitz matrix and the elements of the product can be determined by the elements of the two matrices. Similar to the way we define $\boldsymbol{H}_\theta=\boldsymbol{T}_N(t(\lambda))$ for large $N$, we can define $\boldsymbol{H}_{\rm sr}=\boldsymbol{T}_N(q(\lambda))$ and $\boldsymbol{H}_{\rm rd}=\boldsymbol{T}_N(p(\lambda))$, where $q(\lambda)=\sum_{k=0}^{L_1-1} h_{\rm sr}[k+1]e^{j\lambda k}$ and $p(\lambda)=\sum_{k=0}^{L_2-1} h_{\rm rd}[k+1]e^{j\lambda k}$. Thus, the overall channel matrix is
\begin{align}\label{eqn:freqh}
\boldsymbol{H}_{\rm f}&=\boldsymbol{H}_{\rm rd}\boldsymbol{H}_{\theta}\boldsymbol{H}_{\rm sr}=\alpha_{\rm f}\boldsymbol{T}_N(p(\lambda))\boldsymbol{T}_N(t(\lambda))\boldsymbol{T}_N(q(\lambda))\approx\alpha_{\rm f}\boldsymbol{T}_N(p(\lambda)t(\lambda)q(\lambda)).
\end{align}
$\boldsymbol{H}_{\rm f}$ is also a Toeplitz matrix. Assume the elements in its first column are $h_{\rm f}[k]$ for $k=1,2,\cdots,N$, we have
\begin{align}
h_{\rm f}[k]&=\alpha_{\rm f}\frac{1}{2\pi}\int_{0}^{2\pi}p(\lambda)t(\lambda)q(\lambda)e^{-jk\lambda}{\rm d}\lambda.
\end{align}
The parameters to be estimated are $h_{\rm f}[k]$ for $k=1,2,\cdots,L_{\rm f}$ where $L_{\rm f}=L_1+L_2+L-2$. Assume $\boldsymbol{\xi}_{\rm f}=[h_{\rm f}[1],\cdots,h_{\rm f}[L_{\rm f}]]^T$, our ML estimator can be extended to estimate $\boldsymbol{\xi}_{\rm f}$ by the following. First $h_{\rm f}[1]$ has the same position as $h$ in (\ref{eqn:mlestimator}). Then using $h_{\rm f}[i+1]$ replace $\theta^i$ in (\ref{eqn:mlestimator}). Thus, our ML method can be applied to estimate the overall channel even in the frequency-selective setup. 

The Fisher information for the frequency-selective can be obtained similarly to the flat fading case by using
\begin{align}\label{eqn:freqcrb}
\Gamma_{mn}^{({\rm f})}=\frac{\partial\boldsymbol{\mu}_{{\rm f}}^H}{\partial \xi_{{\rm f}m}^*}\boldsymbol{C}_{\rm f}^{-1}\frac{\partial\boldsymbol{\mu}_{\rm f}}{\partial \xi_{{\rm f}n}}+{\rm tr}\left(\boldsymbol{C}_{\rm f}^{-1}\frac{\partial\boldsymbol{C}_{\rm f}}{\partial \xi_{{\rm f}m}^*}\boldsymbol{C}_{\rm f}^{-1}\frac{\partial\boldsymbol{C}_{\rm f}}{\partial \xi_{{\rm f}n}}\right),
\end{align}
where $\boldsymbol{\mu}_{\rm f}=\boldsymbol{H}_{\rm f}\boldsymbol{x}$ and $\boldsymbol{C}_{\rm f}=\alpha_{\rm f}^2\sigma_{\rm r}^2\boldsymbol{H}_{\rm rd}\boldsymbol{H}_{\theta}\boldsymbol{H}_{\theta}^H\boldsymbol{H}_{\rm rd}^H+\sigma_{\rm d}^2\boldsymbol{I}_N$. The CRBs are given by the diagonal elements of the inverse of the Fisher information matrix $\boldsymbol{\Gamma}^{(\rm f)}$. If desired, a single scalar quantity representing the overall CRB can be computed by find the trace of this matrix: $CRB_{\boldsymbol{\xi}_{\rm f}}=\mathrm{tr}\left((\boldsymbol{\Gamma}^{(\rm f)})^{-1}\right)$.


\section{Multiple relays}\label{sec:multi}
The multi-relay case is also an intuitive extension from the basic one-way relay system. With the analysis for the one-way relay, we can modify the estimation method and CRB analysis. In the multi-relay case, the distance between the source and the destination is fixed. The relays are placed in an equally-spaced manner in series between the source and the destination. Assume there are $M$ relays which satisfy $M(L-1)<N$ (There is a guard time of $L-1$ symbols for each relay). Each relay works in FD mode with AF relay protocol. The relays have their own RSI, and they do not perform estimation or equalization to keep relay complexity low. The estimation and equalization are performed only at the destination. The channel between the $(i-1)$th relay and the $i$th relay is flat fading with coefficients $h_i$ for $i=2,3,\cdots,M$. The channels from the source to the first relay and from the last relay to the destination are $h_1$ and $h_{M+1}$. Channel coefficients $h_i$ for $i=1,\cdots,M+1$ are Gaussian random variables with zero-mean and variance $\sigma_{h}^2$. Each relay has its own RSI channel $h_{{\rm rr}i}$ and power scaling factor $\alpha_i$ which is given by
\begin{align}
\alpha_i^2=\frac{P_{\rm r}}{P_{{\rm s}_{i}}\sigma_{h}^2+P_{\rm r}\sigma_{\rm rr}^2+\sigma_{\rm r}^2},
\end{align} 
where $P_{{\rm s}_{i}}$ is the received power of the desired signal at the $i$th relay. We also assume all the relays have the same average transmit power $P_{\rm r}$ and average RSI power for simplicity. 

The distance between the source and the destination is fixed in our model and $M$ relays are placed in the line between the source and the destination in an equally spaced manner. Assume the distance between the source and the destination is normalized and the corresponding path loss is $K$ dB. Then by using a simplified path loss model\cite{Goldsmithwireless}, the path loss between two relays is 
\begin{align}
K_m dB=K dB+10\gamma\log_{10}(M+1),
\end{align}
where $\gamma$ is the path loss exponent. We incorporate the path loss into $h_i$ which leads to $\sigma_{h}^2=1/K_m$ and $P_{{\rm s}_{i}}=P_{\rm r}K(M+1)^{\gamma}$.

The transmit signal at the $m$th relay is
\begin{align}
\boldsymbol{y}_{m}=\prod_{i=1}^{m}\left(\alpha_i h_i \boldsymbol{H}_{\theta i}\right)\boldsymbol{x}+\sum_{i=1}^{m}\left[\left(\prod_{n=i+1}^{m}\alpha_n h_n \prod_{n=i}^{m}\boldsymbol{H}_{\theta n}\right)\boldsymbol{n}_{{\rm r}i}\right],
\end{align}
where $\boldsymbol{H}_{\theta i}$ is the RSI channel at the $i$th relay defined similarly as $\boldsymbol{H}_{\theta}$ with $\theta_i=\alpha_ih_{{\rm rr}i}$, and $\boldsymbol{n}_{{\rm r}i}$ is the additive Gaussian white noise at the $i$th relay. The received signal at the destination from the $m$th relay is given by
\begin{align}
\boldsymbol{y}_{\rm d}&=h_{M+1}\boldsymbol{y}_{M}+\boldsymbol{n}_{\rm d}=h_{M+1}\prod_{i=1}^{M}\left(\alpha_i h_i \boldsymbol{H}_{\theta i}\right)\boldsymbol{x}+\sum_{i=1}^{M}\left[\left(\prod_{n=i+1}^{m}\alpha_n h_n \prod_{n=i}^{m}\boldsymbol{H}_{\theta n}\right)\boldsymbol{n}_{{\rm r}i}\right]+\boldsymbol{n}_{\rm d}.
\end{align}

Define $\boldsymbol{H}^{(n)}=\prod_{i=n}^{M}\boldsymbol{H}_{\theta i}$ and its corresponding function $t^{(n)}(\lambda)$. By using the property of product of Toeplitz matrix for large $N$, we have
\begin{align}
\boldsymbol{H}^{(n)}=\prod_{i=n}^M \boldsymbol{T}(t_{\theta_i}(\lambda))=\boldsymbol{T}\left(\prod_{i=n}^M t_{\theta_i}(\lambda)\right),
\end{align}
where $t_{\theta_i}(\lambda)$ is defined the same as (\ref{eqn:tfunc}) with $\theta_i$. $\boldsymbol{H}^{(n)}$ is also a Toeplitz matrix and the elements in its first row defined through an inverse Fourier transform 
\begin{align}
h_k^{(n)}=\frac{1}{2\pi}\int_0^{2\pi} t^{(n)}(\lambda)e^{-jk\lambda}d\lambda,
\end{align}
where $t^{(n)}(\lambda)=\prod_{i=n}^M t_{\theta_i}(\lambda)$. Thus we can approximate $\boldsymbol{y}_{\rm d}$ for large $N$ as
\begin{align}\label{eqn:multi_rx}
\boldsymbol{y}_{\rm d}=z_{M}\boldsymbol{H}^{(1)}\boldsymbol{x}+\sum_{i=1}^{M}\left[\left(\prod_{n=i}^{M}\alpha_n h_n \right)/(\alpha_i h_i)\boldsymbol{H}^{(i)}\boldsymbol{n}_{{\rm r}i}\right]+\boldsymbol{n}_{\rm d},
\end{align}
where $z_{M}=h_{M+1}\prod_{i=1}^{M}\alpha_i h_i$. The first row of $\boldsymbol{H}^{(1)}$ is $[1,h_2^{(1)},h_3^{(1)},\cdots,h_{M(L-1)}^{(1)},0,\cdots,0]^T$ which is an $N \times 1$ vector (Assume $M(L-1)<N$). The channel parameters to be estimated is $\boldsymbol{\xi}_M=[z_M,h_2^{(1)},h_3^{(1)},\cdots,h_{M(L-1)}^{(1)}]^T$. The superscript of $h_{m}^{(n)}$ means the overall channel is the channel from the $n$th relay to the last relay while the subscript means the index of the taps of the overall ISI channel, i.e. the index of elements of $\boldsymbol{\xi}_M$. The signal model of (\ref{eqn:multi_rx}) is the same to that of (\ref{eqn:training1}) except additional noise terms. We can also extend our ML estimator to estimate the channel parameters using the same way in the frequency-selective case where $z_M$ and $h_k^{(1)}$ are analogous to $h$ and $\theta^k$ in (\ref{eqn:training1}) respectively. The advantage of estimating the multi-relay channel at the destination rather than at each relay is to keep the relays low complexity with just analog signal processing capability. However, the performance is better if estimation and equalization are performed at each relay, at the cost of complexity.

The CRB for multiple relays can be derived and analyzed similarly to the single relay case. We derive the CRBs for the first two strongest channel taps $z_M$ and $h_2^{(1)}$ which dominate the data detection. The CRBs for other parameters can also be found similarly to (\ref{eqn:freqcrb}). The CRB of $h_2^{(1)}$ is given by
\begin{align}
CRB_{h_2^{(1)}}=\frac{|t^{(M)}(\lambda)|^2|(\sum_{i=1}^{M}|c_i|^2|t_{\theta_i}(\lambda)|^2+1)}{J\left(\sum_{i=1}^{M}|c_i|^2|t_{\theta_i}(\lambda)|^2+1\right)+||\boldsymbol{x}||^2||z_M||^2\left(|t^{(M)}(\lambda)|^2|g^{(M)}(\lambda)|^2-|p^{(M)}(\lambda)|^2\right)},
\end{align}
where $c_i=（\prod_{n=i}^{M}\alpha_n h_n）/(\alpha_i h_i)$ and 
\begin{align}
J=|c_1|^4\frac{1}{2\pi}\int_0^{2\pi}\frac{|t^{(M)}(\lambda)|^2|g^{(M)}(\lambda)|^2}{(\sum_{i=1}^{M}|c_i|^2|t_{\theta_i}(\lambda)|^2+1)^2}.
\end{align}
Define functions $g^{(M)}(\lambda)=\frac{\partial t^{(M)}(\lambda)}{\partial \lambda}$ and $p^{(M)}(\lambda)=\frac{1}{2}\left[(t^{(M)}(\lambda))^*g^{(M)}(\lambda)+t^{(M)}(\lambda)(g^{(M)}(\lambda))^*\right]$.  For $z_M$ we have
\begin{align}
CRB_{z_{M}}=\frac{[|z_M|^2|g^{(M)}(\lambda)|^2|+J(\sum_{i=1}^{M}|c_i|^2|t_{\theta_i}(\lambda)|^2+1)](\sum_{i=1}^{M}|c_i|^2|t_{\theta_i}(\lambda)|^2+1)}{J\left(\sum_{i=1}^{M}|c_i|^2|t_{\theta_i}(\lambda)|^2+1\right)+||\boldsymbol{x}||^2||z_M||^2\left(|t^{(M)}(\lambda)|^2|g^{(M)}(\lambda)|^2-|p^{(M)}(\lambda)|^2\right)}.
\end{align}

We further approximate the CRBs and have simple expressions to find how the number of relays affects the CRBs. From (\ref{eqn:tfunc}) we have $|t_{\theta_i}(\lambda)|\approx 1$ for small $\theta_i$. We also assume the training power is large so that $J \ll ||\boldsymbol{x}||^2$. Therefore, the CRB becomes
\begin{align}
CRB_{h_2^{(1)}}\approx \frac{\sum_{i=1}^{M}|c_i|^2+1}{|z_M|^2||\boldsymbol{x}||^2},\ \ \ CRB_{z_{M}}\approx \frac{\sum_{i=1}^{M}|c_i|^2+1}{||\boldsymbol{x}||^2}.
\end{align}
By plugging in $|c_i|^2=(\alpha_i^2)^{M-i}\prod_{n=i+1}^{M}|h_i|^2$, and $\alpha_i^2=\frac{P_{\rm r}}{P_{\rm r}K(M+1)^\gamma+P_{\rm r}\sigma_{\rm rr}^2+1}$, we have the CRB expressions as a function of $M$.
\begin{align}
CRB_{h_2^{(1)}}&=\frac{\sum_{i=1}^{M}(K(M+1)^\gamma+k_1)^{i-M}(\prod_{n=i+1}^{M}|h_i|^2)^{i-M}+1}{||\boldsymbol{x}||^2(L(M+1)^\gamma+k_1)^{-M}\prod_{n=i+1}^{M}|h_i|^2},\label{eqn:CRBM1}\\
CRB_{z_{M}}&= \frac{\sum_{i=1}^{M}(K(M+1)^\gamma+k_1)^{i-M}(\prod_{n=i+1}^{M}|h_i|^2)^{i-M}+1}{||\boldsymbol{x}||^2}.\label{eqn:CRBM2}
\end{align}
where $k_1=\sigma_{\rm rr}^2+1/P_{\rm r}$. Equation (\ref{eqn:CRBM1}) and (\ref{eqn:CRBM2}) are simple functions of $M$. Intuitively, the estimates of $z_{M}$ will become more inaccurate as the noise goes strong for increasing $M$. However, as the number of relays increases, the RSI for each relay accumulates at the destination, which makes the RSI channel $h_2^{(1)}$ stronger and easier to estimate. Thus there is an optimal $M$ which minimizes the sum MSE of $z_{M}$ and $h_2^{(1)}$. Since $M$ is an integer and is not quite large, the optimal number of relays with respect to the minimum sum CRBs of (\ref{eqn:CRBM1}) and (\ref{eqn:CRBM2}) can be found by searching over $M$.
\section{Numerical results}\label{sec:simulation}
We first simulate the performance of the proposed ML estimator and compare it with the corresponding CRBs. We set $P_r=30$ dB and $\sigma_{\rm rr}^2=-10$ dB. For the channels we set $\sigma_{\rm sr}^2=\sigma_{\rm rd}^2=1$ and the realization of $h_{\rm sr}$ and $h_{\rm rd}$ are drawn from their distributions. The variances of noise at the relay and the destination are set to 1. For each block we estimate the channels and calculate the mean squared error (MSE) which is averaged over multiple independent realizations of the channels. The training length is $N=140$ according to the LTE FDD downlink standard.

\begin{figure}
\begin{minipage}[t]{0.5\textwidth}
\centering
\includegraphics[width=8cm]{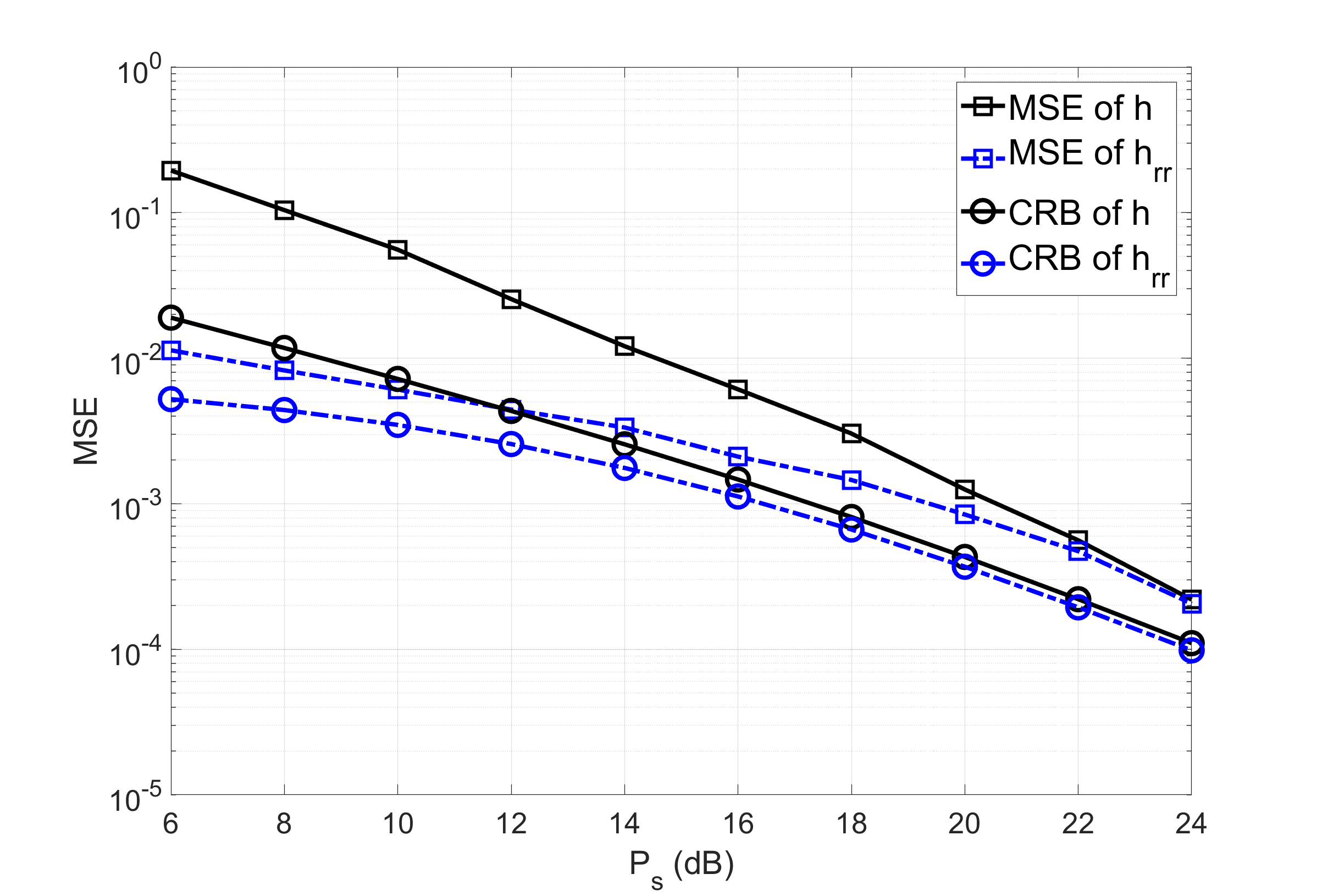}
\caption{Performance of ML estimator compared with CRB.}
\vspace{-7mm}
\label{fig.performance}
\end{minipage}%
\begin{minipage}[t]{0.5\textwidth}
\centering
\includegraphics[width=8cm]{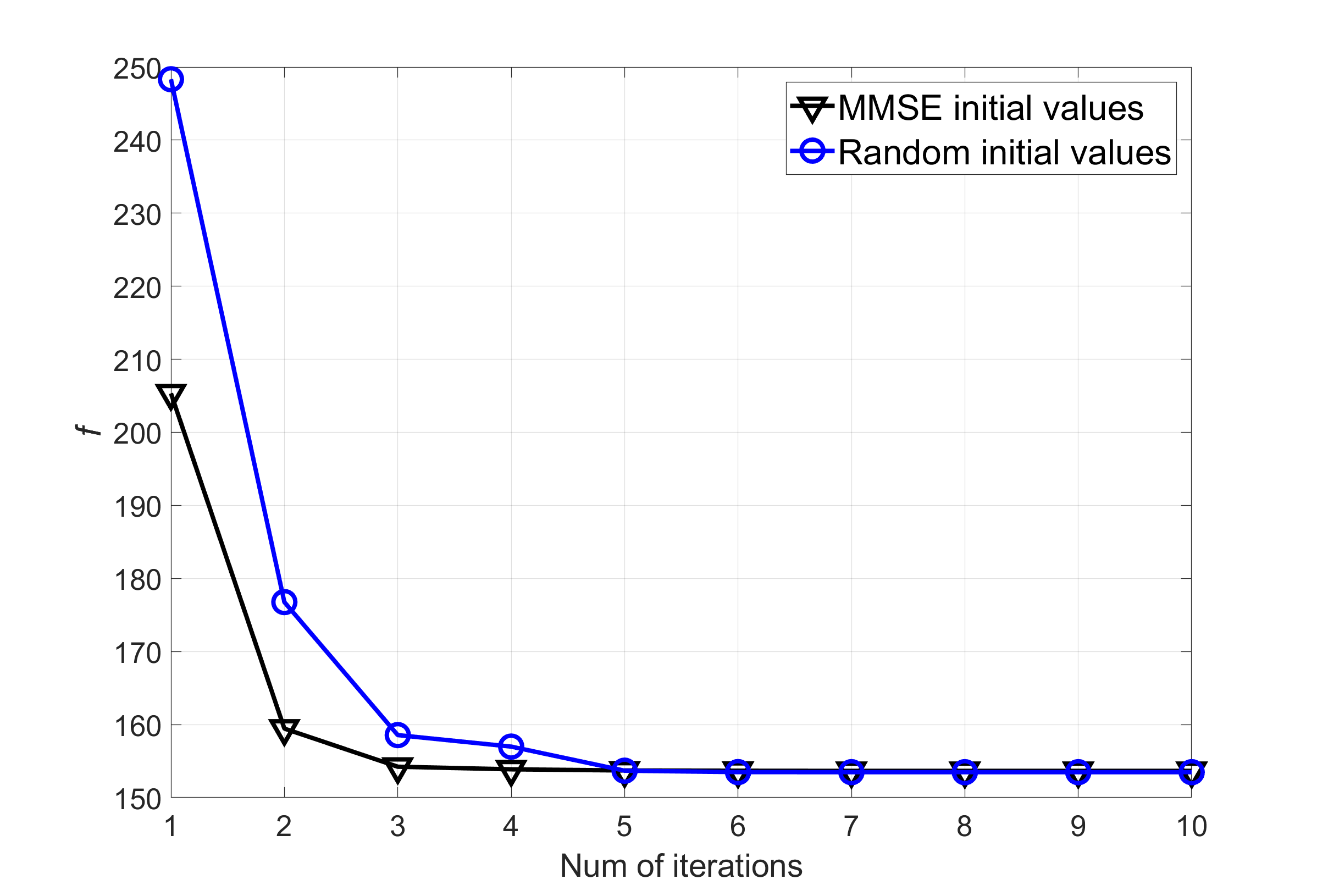}
\caption{Number of iterations to convergence for different initial values.}
\vspace{-7mm}
\label{fig.iteration}
\end{minipage}
\end{figure}

In Figure \ref{fig.performance}, we compare the MSEs of $h$ and $\theta$ to their CRBs. For $h$, we obtain the simulated MSEs of $h_x$ and $h_y$ because our optimization only deals with real numbers. To make a fair comparison with its CRB which is derived for complex numbers, we use the fact that the MSE of $h$ is the sum of the MSEs of its real and imaginary parts. The comparisons for $\theta$ are similar. When $P_{\rm s}$ is small, the RSI dominates the signal, which makes the parameter hard to estimate and results in a large gap between MSE and CRB. Moreover, the colored noise $d\boldsymbol{H}_\theta\boldsymbol{n}_{\rm r} $ also degrades the estimation performance because we use the expectation value of $d$ in the estimation. The effect of colored noise reduces when $P_{\rm s}$ is large. For $\theta$, the MSE does not decrease when $P_{\rm s}$ is less than 10 dB. The MSE for $\theta$ is also affected by the relay power $P_{\rm r}$. It can be seen analytically from (\ref{eqn:gx2}) that when the amplitude of $P_{\rm s}$ is close to that of $P_{\rm r}\sigma_{\rm rr}^2$, the decrease in $\alpha$ is apparent, which leads to a decrease in the CRB.

Figure \ref{fig.iteration} illustrates the convergence speed of the objective function $f$ for different initialization methods, namely, random initialization and MMSE-based initialization. We calculate the average of $f$ in each step for the same $h$ and $\theta$. We observe that with MMSE-based initialization, the objective function converges in 3 iterations while it needs 2 more iterations to converge with random initialization. Thus MMSE-based initialization increases the convergence speed of the algorithm.
\begin{figure}
\begin{minipage}[t]{0.5\textwidth}
\centering
\includegraphics[width=8cm]{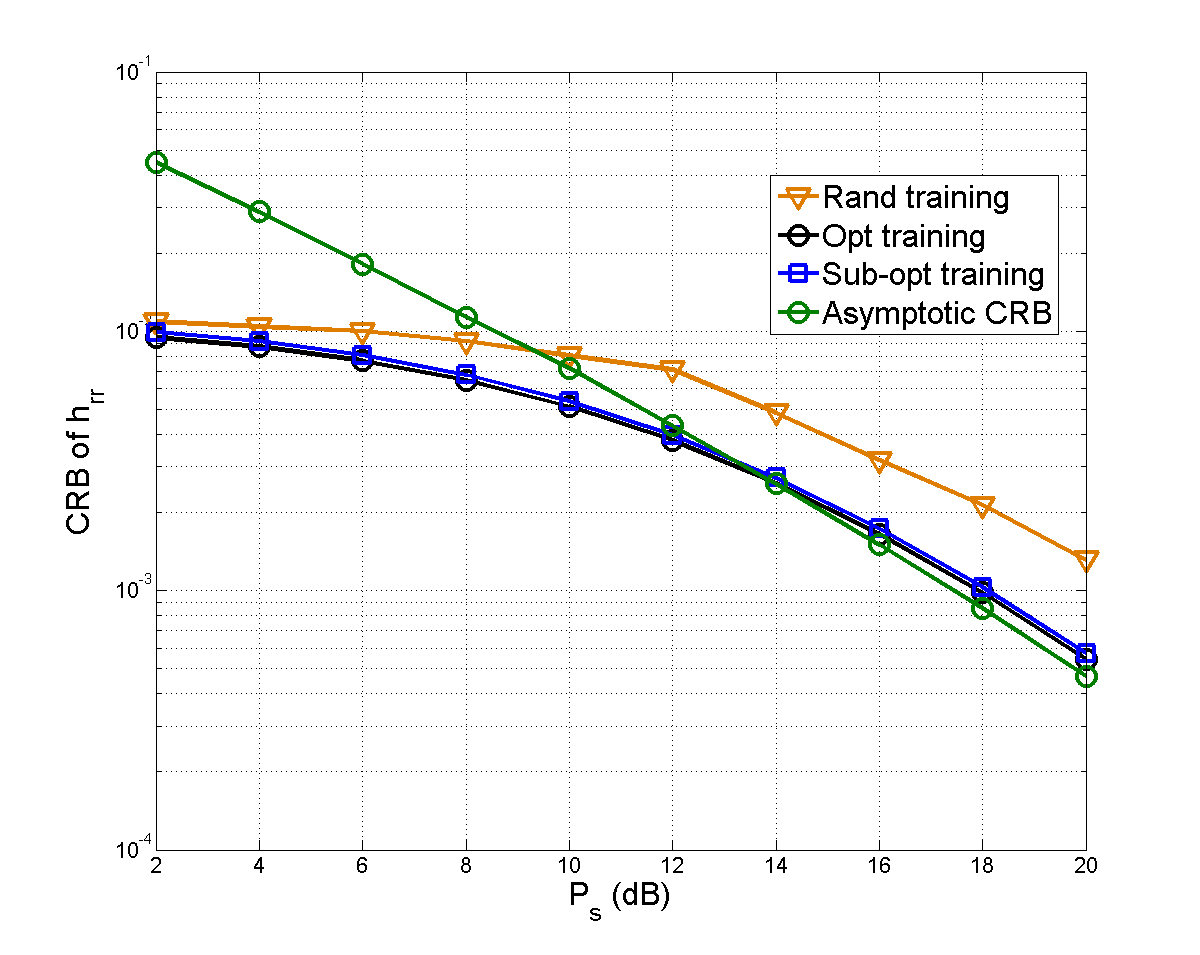}
\caption{Comparison of optimal, approximately optimal, \protect\\and random training sequences.}
\vspace{-7mm}
\label{fig.training_opt}
\end{minipage}%
\begin{minipage}[t]{0.5\textwidth}
\centering
\includegraphics[width=8cm]{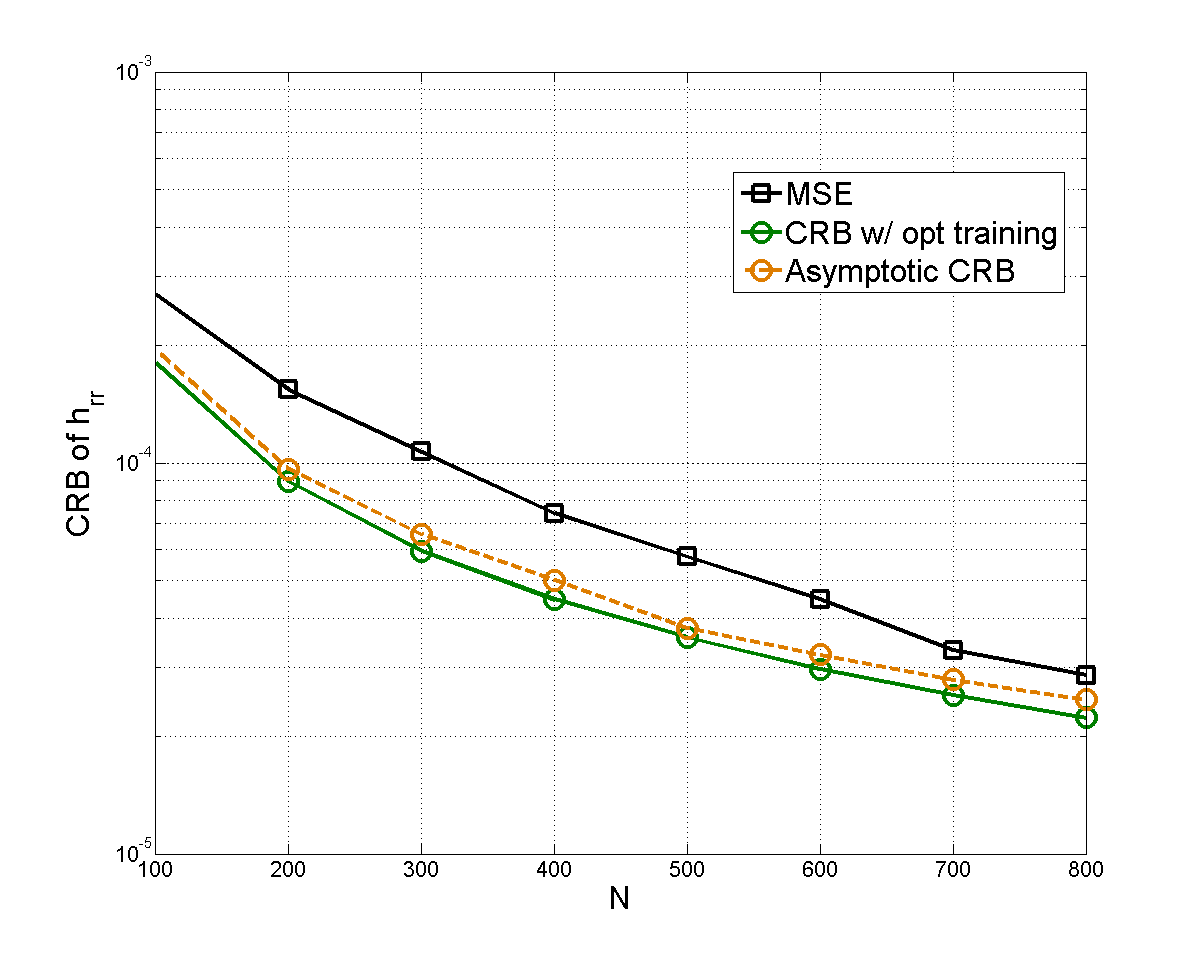}
\caption{Effect of training length $N$ to the CRB.}
\vspace{-7mm}
\label{fig.crb_N}
\end{minipage}
\end{figure}

We compare the CRBs of $\theta$ with different training sequences in Figure \ref{fig.training_opt}. We generate a training sequence which consists of i.i.d Bernoulli symbols which are random $+1$ and $-1$ with equal probability to compare with the optimal training sequence. The optimal and approximately optimal curves are almost overlapped. We observe that for different $\theta$ values, the roots calculated from the 8th order equation do not fall in the interval $[(1-|\theta|)^2,(1+|\theta|)^2]$ discussed in Section \ref{sec:szego}. Thus the optimal solution is on the boundary values which is consistent with the approximately optimal solution. Therefore the simulation results of the optimal and the approximately optimal cases are very close. Figure \ref{fig.training_opt} also shows that the optimal training sequences save approximately 3dB in power compared to the case of random training sequences.

Figure \ref{fig.crb_N} shows the influences of training length $N$ on the simulated CRB and the CRB calculated asymptotically, with the optimal training sequence. As $N$ increases, the MSE gradually gets close to the CRB as we expect, since the ML estimator is asymptotically efficient (when $N$ goes to infinity)\cite{MK1993} which shows the estimate of the RSI channel gets more accurate. The asymptotic CRB from (\ref{eqn:crb_asymp1}) is an approximation of the CRB from (\ref{eqn:crbtheta}) when the training length goes to infinity, and has a closed-form expression which can be efficiently calculated and analyzed. The small gap in the simulation shows the accuracy of the approximation.
 	
\begin{figure}
\begin{minipage}[t]{0.5\textwidth}
\centering
\includegraphics[width=8cm]{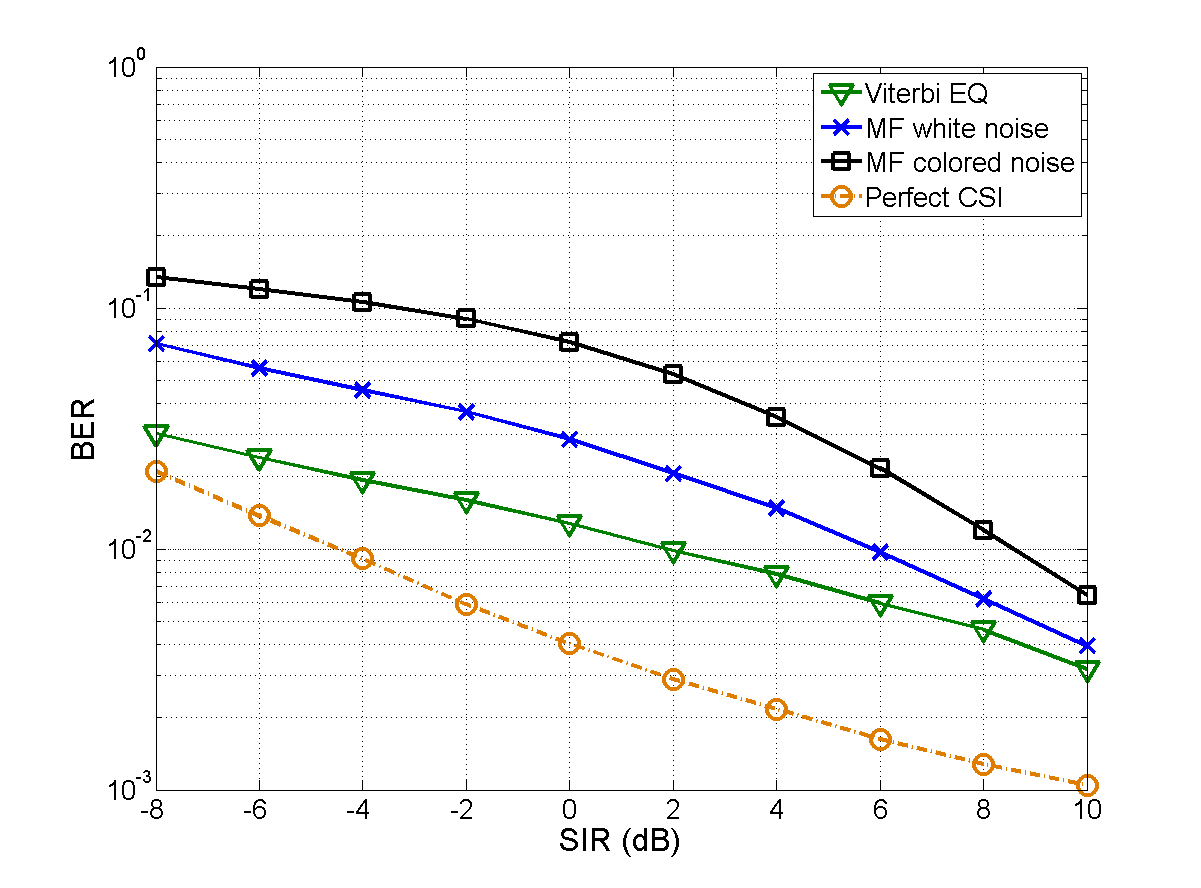}
\caption{BER comparison of different detectors.}
\vspace{-7mm}
\label{fig.ber}
\end{minipage}%
\begin{minipage}[t]{0.5\textwidth}
\centering
\includegraphics[width=8cm]{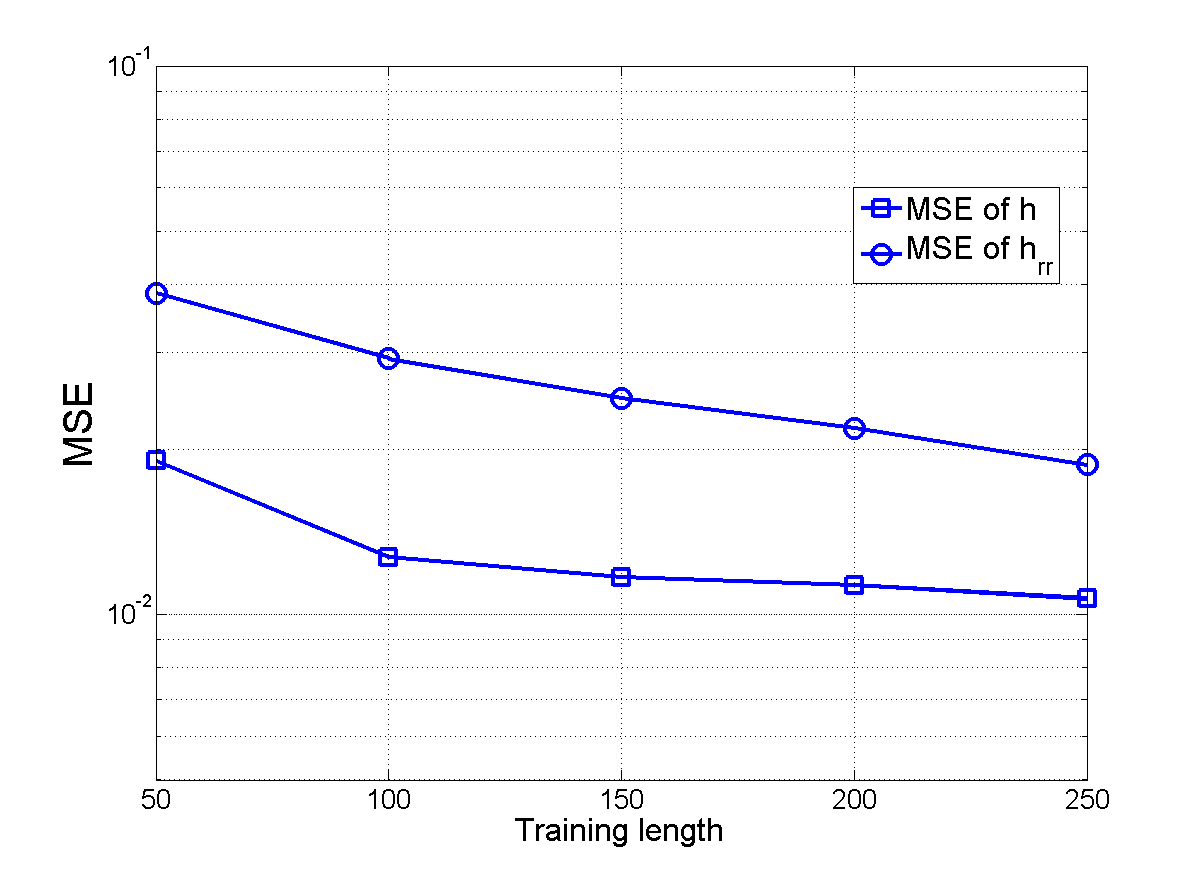}
\caption{MSE with increasing $N$ in frequency-selective case.}
\vspace{-7mm}
\label{fig.freqN}
\end{minipage}
\end{figure}

In Figure \ref{fig.ber}, we compare the performance of different detectors including the Viterbi equalizer and matched filter (MF) detector with channel tap length $L = 3$. The Viterbi equalizer we use is the standard one for ISI channel\cite{Goldsmithwireless}. However, in our system the ISI is caused by the RSI and forms a channel with taps $[h,h\theta,h\theta^2]^T$ which can be obtained from the estimates of $h$ and $\theta$. Thus, the standard Viterbi equalizer can be applied for RSI mitigation. On the other hand, the MF detects the signal by multiplying the received signal by  the strongest tap of the ISI channel which is $h$. There are also two cases for MF detector. In one case, MF detector is directly used to the received signal which has colored noise. The other case is obtained by first applying a noise whitening filter to the received signal and then doing MF detection. So the noise is whitened in this case. From the perspective of how much CSI is needed, the former case only needs $h$ while the latter needs both $h$ and $\theta$. Figure 6 shows the Viterbi equalizer outperforms any MF detector since the equalizer cancels the RSI while MF treats the RSI as noise. For the two MF detectors, the one that whitens the noise has better performance which comes from the noise whitening filter by using the CSI of $\theta$. The fact that canceling the RSI and whitening the noise lead to better performance illustrates the benefits of estimating the RSI channel $\theta$.

\begin{figure}
    \centering
    \includegraphics[width=8cm]{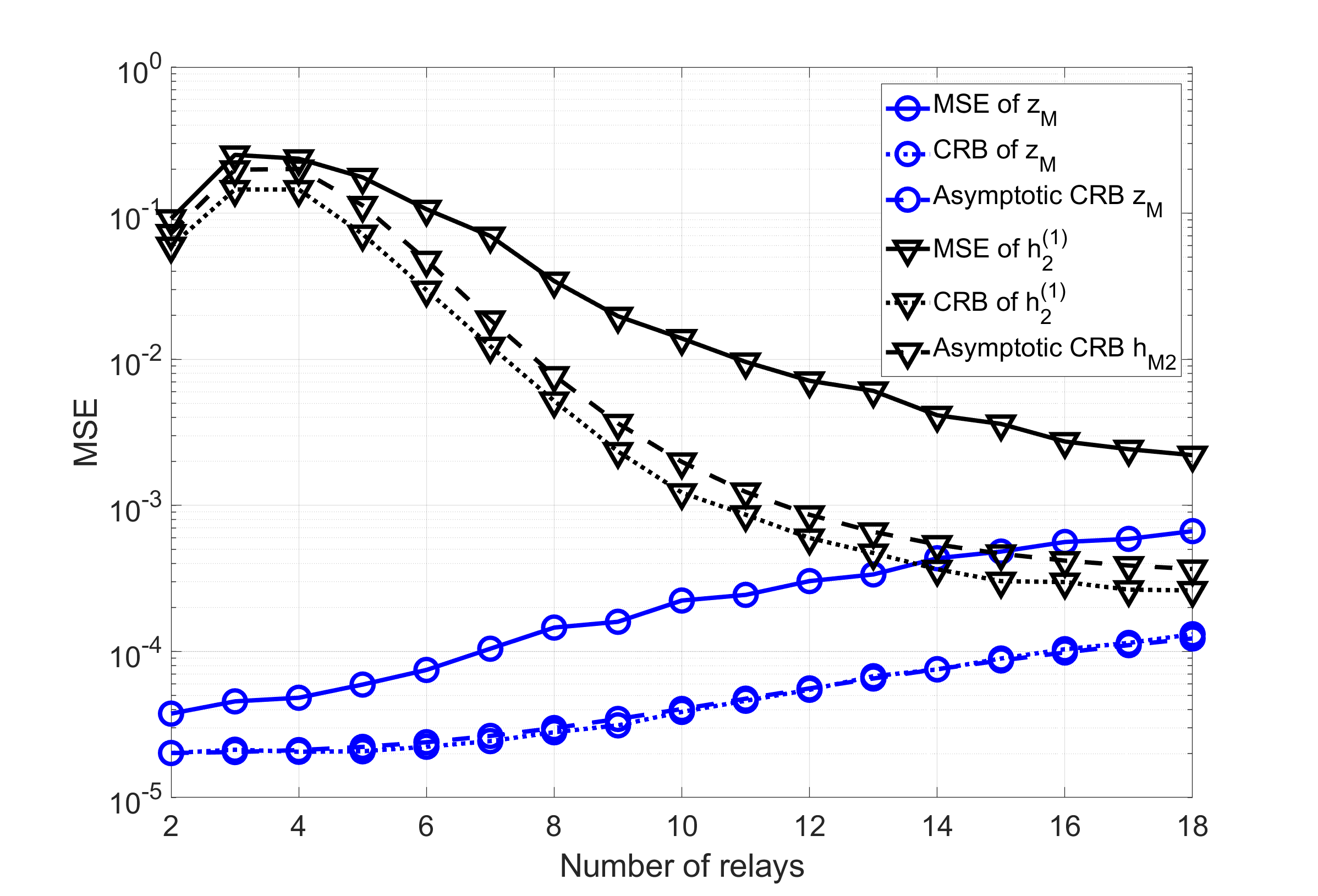}
    \vspace{-4mm}
    \caption{CRB for multiple relays.}
    \label{fig.multir}
    \vspace{-8mm}
\end{figure}

Figures \ref{fig.freqN} and \ref{fig.multir} show the MSE of the two extensions. Note that the estimator are derived by using $\boldsymbol{H}_{\rm f}$ in (\ref{eqn:freqh}) and $\boldsymbol{H}^{(1)}$ in (\ref{eqn:multi_rx}) which are the approximation of the exact channels for frequency-selective case and multi-relay case respectively. The MSE is calculated by comparing the estimates of the approximation to the exact channels. Figure \ref{fig.freqN} shows that in the frequency-selective case, the MSE reduces with the training length $N$ increasing, implying that the asymptotic approximation $\boldsymbol{H}_{\rm f}$ gets closer to the exact channels. The reducing MSE shows the accuracy of the approximation. 


Figure \ref{fig.multir} shows the MSEs of $z_M$ and $h_2^{(1)}$ compared with their CRBs in the multi-relay case. Specifically, the total path loss between the source and the relay is $K=-60$ dB and path-loss exponent is $\gamma=3.71$ for the outdoor environment. As $M$ increases, the MSE of $z_M$ increases because more noise and interference are added. On the other hand, the MSE of $h_2^{(1)}$ decreases, since the RSI channel is easier to estimate as the RSI gets stronger. The asymptotic CRBs derived by (\ref{eqn:CRBM1}) and (\ref{eqn:CRBM2}) are close to the simulated CRBs. Since $M$ is an integer and not large, one can search over the best $M$ by using (\ref{eqn:CRBM1}) and (\ref{eqn:CRBM2}).
\vspace{-5mm}
\section{Conclusion}\label{sec:conclusion}
We propose an ML channel estimator in FD relays to estimate the end-to-end channel as well as the RSI channel at the destination. The log-likelihood function is maximized through the BFGS algorithm. The algorithm is initialized by a linear MMSE estimator to prevent local minima and increase the convergence speed. The corresponding CRBs are derived to evaluate the accuracy of the estimates. By using asymptotic properties of Toeplitz matrices, we show that the optimal training sequence is a sinusoid. To find the frequency, we minimize the CRBs and propose the corresponding optimal training sequence and a practical approximately optimal training sequence. Extensions of our estimation method to frequency-selective and multi-relay case are also considered.

\newpage
\section*{Appendix \rom{1} \\Mean and Covariance matrix of $p(\boldsymbol{y}|h,\theta)$}\label{sec:App1}
In (\ref{eqn:trainingdata}), $h$ and $\theta$ are parameters of interest and $d$ is the nuisance parameter. The likelihood function $p(\boldsymbol{y}|h,\theta)$ is obtained through integrating $p(\boldsymbol{y}|h,\theta,d)$ with respect to $d$\cite{MK1993}, 
\begin{align}
p(\boldsymbol{y}|h,\theta)=\int p(\boldsymbol{y}|h,\theta,d)p(d)\mathrm{d}d.
\end{align}
Since $p(\boldsymbol{y}|h,\theta,d)$ and $p(d)$ are Gaussian distributed, it can be shown that the distribution of $p(\boldsymbol{y}|h,\theta)$ is also Gaussian. Denoting the mean of $\boldsymbol{y}$ given $h$ and $\theta$ to be $E[\boldsymbol{y}|h,\theta]$ and the covariance matrix as $V[\boldsymbol{y}|h,\theta]$. It can be shown that
\begin{align}
E[\boldsymbol{y}|h,\theta]=E_d[E_{\boldsymbol{y}}[\boldsymbol{y}|h,\theta,d]], \ \ \ V[\boldsymbol{y}|h,\theta]=V_d[E_{\boldsymbol{y}}[\boldsymbol{y}|h,\theta,d]]+E_d[V_{\boldsymbol{y}}[\boldsymbol{y}|h,\theta,d]].
\end{align}
Since we know $p(\boldsymbol{y}|h,\theta,d)$ is a Gaussian distribution with mean $\boldsymbol{\mu}$ and co-variance matrix $\boldsymbol{C}$, then it is straight forward to get
\begin{align}
E_{\boldsymbol{y}}[\boldsymbol{y}|h,\theta,d]=h\boldsymbol{H}_\theta,\ \ \ \ V_{\boldsymbol{y}}[\boldsymbol{y}|h,\theta,d]=|d|^2\sigma_{\rm r}^2\boldsymbol{H}_\theta\boldsymbol{H}_\theta^H+\sigma_{\rm d}^2\boldsymbol{I}_N.
\end{align}
The distribution of $p(d)$ is also Gaussian with zero mean and variance $\alpha^2$, thus
\begin{align}
E_d[E_{\boldsymbol{y}}[\boldsymbol{y}|h,\theta,d]]&=E_d[h\boldsymbol{H}_\theta]=h\boldsymbol{H}_\theta, \ \ V_d[E_{\boldsymbol{y}}[\boldsymbol{y}|h,\theta,d]]=V_d[h\boldsymbol{H}_\theta]=\boldsymbol{0}_{N\times N},\\
E_d[V_{\boldsymbol{y}}[\boldsymbol{y}|h,\theta,d]]&=E_d[|d|^2\sigma_{\rm r}^2\boldsymbol{H}_\theta\boldsymbol{H}_\theta^H+\sigma_{\rm d}^2\boldsymbol{I}_N]=\alpha^2\sigma_{\rm r}^2\boldsymbol{H}_\theta\boldsymbol{H}_\theta^H+\sigma_{\rm d}^2\boldsymbol{I}_N.
\end{align}
Therefore, we can obtain the mean and covariance matrix of the Gaussian distribution $p(\boldsymbol{y}|h,\theta)$,
\begin{align}
\boldsymbol{\mu}=E[\boldsymbol{y}|h,\theta]&=h\boldsymbol{H}_\theta, \ \ \boldsymbol{C}=V[\boldsymbol{y}|h,\theta]=\alpha^2\sigma_{\rm r}^2\boldsymbol{H}_\theta\boldsymbol{H}_\theta^H+\sigma_{\rm d}^2\boldsymbol{I}_N.
\end{align}
\vspace{-5mm}
\section*{Appendix \rom{2} \\Unknown Delay $\tau_0$ at the Relay}
We first begin considering our discrete signal model under unknown $\tau_0\in\mathbb{R}^+$. When the desired received signal and the RSI signal are not synchronized, we jointly estimate $\tau_0$ and the channels. Assume that the analog transmitted signal, the received signal at the relay, and the received signal at the destination are $x(t)$, $y_{\rm r}(t)$, and $y_{\rm d}(t)$ respectively. The source-relay, relay-destination, and RSI channels are $h_{\rm sr}(\tau)=h_{\rm sr}\delta(\tau)$, $h_{\rm rd}(\tau)=h_{\rm rd}\delta(\tau)$ and $h_{\rm rr}(\tau)=\theta\delta(\tau)$ where $h_{\rm sr}$, $h_{\rm rd}$ and $\theta$ are the channel coefficients and the variable $\tau$ is the delay of the channels. The three channels are assumed to be time-invariant and flat fading. Let $s[n]$ be the sequence that the source transmitted. The source transmit signal is $x(t)=\sum_{n=-\infty}^{+\infty}s[n]p^{\rm tr}(t-nT_s)$ where $T_s$ is the symbol duration and $p^{\rm tr}(t)$ is a pulse shaping filter. The relay forwards the superimposed signal of the desired received signal and the RSI signal to the destination. By ignoring the noise, we have
\vspace{-1mm}
\begin{align}
y_{\rm r}(t)=(h_{\rm sr}*x)(t)+(h_{\rm rr}*y_{\rm r})(t-\tau_0),\ \ \ y_{\rm d}(t)=(h_{\rm rd}*y_{\rm r}*p^{\rm rec})(t),
\end{align}
where $p^{\rm rec}(t)$ is the matched filter at the destination. The notation $*$ stands for convolution. The delayed version of $y_{\rm r}(t)$ is $y_{\rm r}(t-\tau_0)=(h_{\rm sr}*x)(t-\tau_0)+(h_{\rm rr}*y_{\rm r})(t-2\tau_0)$. By substituting $y_{\rm r}(t-\tau_0)$ and $x(t)$ into $y_{\rm d}(t)$, we have
\vspace{-1mm}
\begin{align}\label{eqn:continuousmodel}
y_{\rm d}(t)=h\sum_{n=-\infty}^{+\infty}s[n]\left(\sum_{l=0}^{L-1}\theta^l p(t-nT_s-l\tau_0)\right)
\end{align}
where $L$ is the effective length of the overall channel impulse response, $|\theta|^L\approx0$ by the finite energy assumption, coefficient $h=h_{\rm rd}h_{\rm sr}$, and $p(t)=p^{\rm tr}(t)*p^{\rm rec}(t)$ is a raised-cosine filter. At the destination, we sample the received signal with the same sampling rate $1/T_s$. The sampled signal is
\vspace{-1mm}
\begin{align}
y_{\rm d}(kT_s)=h\sum_{n=-\infty}^{+\infty}s[n]\left(\sum_{l=0}^{L-1}\theta^l p(kT_s-nT_s-l\tau_0)\right).
\end{align}
We can obtain a discrete signal model
\vspace{-1mm}
\begin{align}
y_{\rm d}(kT_s)=y_{\rm d}[k]=s[n]*h[k-n],
\end{align}
where
\vspace{-5mm}
\begin{align}\label{eqn:hkt1}
h[k]=\sum_{l=0}^{L-1}h\theta^l p(kT_s-l\tau_0)\triangleq h g[k].
\end{align}
Note that since $\tau_0$ is not an integer multiple of $T_s$, the delayed signal and the source-relay signal are not synchronized. The sampling point of the overlapped signal of the two signals is not exactly the zero positions of the raised-cosine filter. Therefore, the resulting discrete-time channel model has non-zero coefficients given by (\ref{eqn:hkt1}). Define an $N$ by $N$ Toeplitz matrix as $\boldsymbol{H}[\boldsymbol{a}]$ whose first column is $[a[0],a[1],\cdots,a[M-1],0,\cdots,0]^T$ and first row is $[a[0],0,\cdots,0]$ where $\boldsymbol{a}=[a[0],a[1],\cdots,a[M]]^T$, $M\leq N$. Therefore, the channel matrix is $h\boldsymbol{H}[\boldsymbol{g}]$ with $\boldsymbol{g}=[g[0],g[1],\cdots,g[N-1]]^T$. The destination received signal in vector form is
\vspace{-1mm}
\begin{align}
\boldsymbol{y}=h\boldsymbol{H}[\boldsymbol{g}]\boldsymbol{x}+d\boldsymbol{H}[\boldsymbol{g}]\boldsymbol{n}_{\rm r}+\boldsymbol{n}_{\rm d}.
\end{align}

\textbf{\underline{ML estimator:}} We can find the mean and covariance matrix of $\boldsymbol{y}$ as
\vspace{-1mm}
\begin{align}
\boldsymbol{\mu}=h\boldsymbol{H}[\boldsymbol{g}],\ \ \ \boldsymbol{C}=\alpha^2\sigma_{\rm r}^2\boldsymbol{H}[\boldsymbol{g}]\boldsymbol{H}^H[\boldsymbol{g}]+\sigma_{\rm d}^2\boldsymbol{I}_N.
\end{align}
The ML estimator is given by
\vspace{-1mm}
\begin{align}\label{eqn:mlestimator1}
\{\hat{h},\hat{\theta},\tau_0\}=\argmin_{h,\theta,\tau_0}&\big\{\log|\boldsymbol{C}|+(\boldsymbol{y}-\boldsymbol{\mu})^H\boldsymbol{C}^{-1}(\boldsymbol{y}-\boldsymbol{\mu})\big\}.
\end{align}
We can eliminate $h$ and find the derivatives of the log-likelihood function with respect to $\theta$ by the same way as the case with known $\tau_0$. For the unknown $\tau_0$, we need to find the derivatives of the log-likelihood function with respect to it as follows.

Define $b[k]$ and $q[k]$ as the derivatives of $g[k]$ with respect to $\theta$ and $\tau_0$ respectively,
\vspace{-1mm}
\begin{align}
b[k]=\sum_{l=0}^{L-1}l\theta^{l-1} p(kT_s-l\tau_0), \ \ \ q[k]=-\sum_{l=0}^{L-1}l\theta^{l} p'(kT_s-l\tau_0)
\end{align}
where $p'(t)$ is the derivative of $p(t)$. The derivative of $\boldsymbol{C}$ with respect to $\tau_0$ is
\vspace{-1mm}
\begin{align}
\boldsymbol{B}_{\tau_0}=\frac{\partial \boldsymbol{C}}{\partial \tau_0}=\alpha^2\sigma_{\rm r}^2\left(\boldsymbol{H}[\boldsymbol{g}]\boldsymbol{H}^H[\boldsymbol{q}]+\boldsymbol{H}^H[\boldsymbol{g}]\boldsymbol{H}[\boldsymbol{q}]\right).
\end{align}
Therefore, we have
\vspace{-1mm}
\begin{align}
\frac{\partial f}{\partial \tau_0}=-{\rm tr}\left(\boldsymbol{C}^{-1}\boldsymbol{B}_{\tau_0}\right)+(\boldsymbol{y}-\boldsymbol{\mu})^H\boldsymbol{C}^{-1}\boldsymbol{B}_{\tau_0}\boldsymbol{C}^{-1}(\boldsymbol{y}-\boldsymbol{\mu})+2{\rm Re}\left[(\boldsymbol{y}-\boldsymbol{\mu})^H\boldsymbol{C}^{-1}h\boldsymbol{H}[\boldsymbol{q}]\boldsymbol{x}\right].
\end{align}
With the gradients of $\tau_0$ and $\theta$, the ML estimator can be solved by the BFGS algorithm. 

\textbf{\underline{CRBs:}} Let $\boldsymbol{\xi}=[h\ \theta\ \tau_0]^T$ be the vector of unknown parameters. The Fisher information of the parameters is given by the following.
\vspace{-1mm}
\begin{align}
\Gamma_{11}&=\boldsymbol{x}^H \boldsymbol{H}^H[\boldsymbol{g}]\boldsymbol{C}^{-1}\boldsymbol{H}[\boldsymbol{g}]\boldsymbol{x},\\
\Gamma_{22}&=|h|^2\boldsymbol{x}^H\boldsymbol{H}^H[\boldsymbol{b}]\boldsymbol{C}^{-1}\boldsymbol{H}[\boldsymbol{b}]\boldsymbol{x}+\alpha^4\sigma_{\rm r}^4{\rm tr}\left(\boldsymbol{C}^{-1}\boldsymbol{H}[\boldsymbol{g}]\boldsymbol{H}^H[\boldsymbol{b}]\boldsymbol{C}^{-1}\boldsymbol{H}[\boldsymbol{b}]\boldsymbol{H}^H[\boldsymbol{g}]\right),\\
\Gamma_{33}&=|h|^2\boldsymbol{x}^H \boldsymbol{H}^H[\boldsymbol{q}]\boldsymbol{C}^{-1}\boldsymbol{H}[\boldsymbol{q}]\boldsymbol{x}+{\rm tr}\left(\boldsymbol{C}^{-1}\boldsymbol{B}_{\tau_0}\boldsymbol{C}^{-1}\boldsymbol{B}_{\tau_0}\right).
\end{align}
Similarly, $\Gamma_{12}$, $\Gamma_{13}$, and $\Gamma_{23}$ can be found. The CRBs are given by the trace of the inverse of the Fisher information matrix.

\textbf{\underline{Updated asymptotic CRB analysis:}} Next we explain how to update our CRB analysis with unknown $\tau_0$. The key idea in the analysis is that the Toeplitz channel matrix behaves as a circulant matrix asymptotically when the size of it goes to infinity and we can find the eigenvalues and corresponding eigenvectors through function $t(\lambda)$ which characterizes the asymptotic circulant matrix. To modify the analysis, we need to find a new function $t_{\tau_0}(\lambda)=\sum_{k=-\infty}^{+\infty}h[k]e^{j\lambda k}$ which is a discrete-time Fourier transform (DTFT) of $h[k]$ in closed-form. When the delay $l\tau_0$ in $h[k]$ is not an integer multiple of $T_s$, the DTFT with a time shift cannot be applied directly. We use the relationship of the continuous signal $h_{\rm c}(t)$ of which $h[n]$ are the samples, the impulse train $h_{\rm p}(t)$ with amplitudes corresponding to the samples of $h_{\rm c}(t)$, and the discrete samples $h[n]$ \cite{Oppenheimsignal}. We can find the DTFT of $h[n]$ as 
\vspace{-1mm}
\begin{align}
H_{\rm d}(e^{j\Omega})=\frac{1}{T_s}\sum_{l=0}^{L-1}h\theta^l P_{\rm c}\left(j\Omega/T_s\right)e^{-j\Omega l\tau_0/T_s},
\end{align}
where $P_{\rm c}(j\omega)$ is the continuous Fourier transform of $h_{\rm c}(t)$, $\Omega$ is the frequency variable with period $2\pi/T_s$. We have $t_{\tau_0}(\lambda)=H_{\rm d}(e^{j\lambda})$. Therefore, our CRB analysis based on the closed-form expression of $t(\lambda)$ also works for $t_{\tau_0}(\lambda)$.

\vspace{-5mm}
\section*{Appendix \rom{3} \\The Gradients used in the BFGS algorithm}\label{sec:App3}
Now we derive the gradients of $f$ with respect to $\theta_x$ and $\theta_y$ which are used in the BFGS algorithm. The gradients for both real and imaginary parts are needed as inputs of the algorithm. For $\theta$, we first obtain the derivative of $\boldsymbol{H}_\theta$ with respect to $\theta$, denoted as $\boldsymbol{B}_{\theta}$, which is also an $N \times N$ Toeplitz matrix with first column $[0,1,2\theta,\cdots,(L-1)\theta^{L-2},0,\cdots,0]^T$ and first row $\boldsymbol{0}_{N\times 1}^T$.

Both $\boldsymbol{C}$ and $\boldsymbol{\mu}$ contain $\theta$, therefore there are three terms in its gradient. We have
\begin{align}
&\nabla f_{\theta_x}=\mathrm{tr}\left(\alpha^2\sigma_{\rm r}^2\boldsymbol{C}^{-1}(\boldsymbol{B}_\theta\boldsymbol{H}_\theta^H+\boldsymbol{H}_\theta\boldsymbol{B}_\theta^H)\right)-2\mathrm{Re}\left[(\boldsymbol{y}-\boldsymbol{\mu})^H\boldsymbol{C}^{-1}h\boldsymbol{B}_\theta\boldsymbol{x}\right]\nonumber\\
&-\alpha^2\sigma_{\rm r}^2(\boldsymbol{y}-\boldsymbol{\mu})^H\boldsymbol{C}^{-1}(\boldsymbol{B}_\theta\boldsymbol{H}_\theta^H+\boldsymbol{H}_\theta\boldsymbol{B}_\theta^H)\boldsymbol{C}^{-1}(\boldsymbol{y}-\boldsymbol{\mu}),\label{eqn:grad1}\\
&\nabla f_{\theta_y}=\mathrm{tr}\left(j\alpha^2\sigma_{\rm r}^2\boldsymbol{C}^{-1}(\boldsymbol{B}_\theta\boldsymbol{H}_\theta^H-\boldsymbol{H}_\theta\boldsymbol{B}_\theta^H)\right)-2\mathrm{Re}\left[(\boldsymbol{y}-\boldsymbol{\mu})^H\boldsymbol{C}^{-1}jh\boldsymbol{B}_\theta\boldsymbol{x}\right]\nonumber\\
&-j\alpha^2\sigma_{\rm r}^2(\boldsymbol{y}-\boldsymbol{\mu})^H\boldsymbol{C}^{-1}(\boldsymbol{B}_\theta\boldsymbol{H}_\theta^H-\boldsymbol{H}_\theta\boldsymbol{B}_\theta^H)\boldsymbol{C}^{-1}(\boldsymbol{y}-\boldsymbol{\mu}).\label{eqn:grad2}
\end{align}

\bibliographystyle{IEEEtran}
\bibliography{twrc_fundamental}

\begin{thebibliography}{10}
\providecommand{\url}[1]{#1}
\csname url@samestyle\endcsname
\providecommand{\newblock}{\relax}
\providecommand{\bibinfo}[2]{#2}
\providecommand{\BIBentrySTDinterwordspacing}{\spaceskip=0pt\relax}
\providecommand{\BIBentryALTinterwordstretchfactor}{4}
\providecommand{\BIBentryALTinterwordspacing}{\spaceskip=\fontdimen2\font plus
\BIBentryALTinterwordstretchfactor\fontdimen3\font minus
  \fontdimen4\font\relax}
\providecommand{\BIBforeignlanguage}[2]{{%
\expandafter\ifx\csname l@#1\endcsname\relax
\typeout{** WARNING: IEEEtran.bst: No hyphenation pattern has been}%
\typeout{** loaded for the language `#1'. Using the pattern for}%
\typeout{** the default language instead.}%
\else
\language=\csname l@#1\endcsname
\fi
#2}}
\providecommand{\BIBdecl}{\relax}
\BIBdecl

\bibitem{Heino15full}
M.~Heino, D.~Korpi, T.~Huusari, E.~Antonio-Rodriguez, S.~Venkatasubramanian,
  T.~Riihonen, L.~Anttila, C.~Icheln, K.~Haneda, and R.~Wichman, ``Recent
  advances in antenna design and interference cancellation algorithms for
  in-band full duplex relays,'' \emph{IEEE Communications Magazine}, vol.~53,
  no.~5, pp. 91--101, Oct. 2015.

\bibitem{hong2014sic5g}
S.-K. Hong, J.~Brand, J.~Choi, M.~Jain, J.~Mehlman, S.~Katti, and P.~Levis,
  ``Applications of self-interference cancellation in 5{G} and beyond,''
  \emph{IEEE Communications Magazine}, vol.~52, no.~2, pp. 114--121, Oct. 2014.

\bibitem{sabharwal2014band}
A.~Sabharwal, P.~Schniter, D.~Guo, D.~W. Bliss, S.~Rangarajan, and R.~Wichman,
  ``In-band full-duplex wireless: Challenges and opportunities,'' \emph{IEEE
  Journal on Selected Areas in Communications}, vol.~32, no.~9, pp. 1637--1652,
  Sep. 2014.

\bibitem{ma2009new}
J.~Ma, G.~Y. Li, J.~Zhang, T.~Kuze, and H.~Iura, ``A new coupling channel
  estimator for cross-talk cancellation at wireless relay stations,'' in
  \emph{Proc. IEEE Global Telecommun. Conf.}, Oct. 2009, pp. 1--6.

\bibitem{masmoudi2015mlsic}
A.~Masmoudi and T.~Le-Ngoc, ``A maximum-likelihood channel estimator for
  self-interference cancellation in full-duplex systems,'' \emph{IEEE
  Transactions on Vehicular Technology}, vol.~65, no.~7, pp. 5122--5132, Oct.
  2016.

\bibitem{Koohian2015estifull}
A.~Koohian, H.~Mehrpouyan, M.~Ahmadian, and M.~Azarbad, ``Bandwidth efficient
  channel estimation for full duplex communication systems,'' in \emph{Proc.
  IEEE ICC}, Oct. 2015, pp. 4710--4714.

\bibitem{riihonen2011mitigation}
T.~Riihonen, S.~Werner, and R.~Wichman, ``Mitigation of loopback
  self-interference in full-duplex {MIMO} relays,'' \emph{IEEE Transactions on
  Signal Processing}, vol.~59, no.~12, pp. 5983--5993, Dec. 2011.

\bibitem{Duarte12full}
M.~Duarte, C.~Dick, and A.~Sabharwal, ``Experiment-driven characterization of
  full-duplex wireless systems,'' \emph{IEEE Transactions on Wireless
  Communications}, vol.~11, no.~12, pp. 4296--4307, Dec. 2012.

\bibitem{riihonen2010residual}
T.~Riihonen, S.~Werner, and R.~Wichman, ``Residual self-interference in
  full-duplex {MIMO} relays after null-space projection and cancellation,'' in
  \emph{Proc. IEEE 44th Asilomar Conf. Signals, Syst. Comput.}, Nov. 2010, pp.
  653--657.

\bibitem{Kim2012full}
T.~M. Kim and A.~Paulraj, ``Outage probability of amplify-and-forward
  cooperation with full duplex relay,'' in \emph{Proc. IEEE WCNC}, Oct. 2012,
  pp. 75--79.

\bibitem{jimenez2014per}
L.~Jimenez~Rodriguez, N.~H. Tran, and T.~Le-Ngoc, ``Performance of full-duplex
  af relaying in the presence of residual self-interference,'' \emph{IEEE
  Journal on Selected Areas in Communications}, vol.~32, no.~9, pp. 1752--1764,
  Jun. 2014.

\bibitem{jimenez2014power}
------, ``Optimal power allocation and capacity of full-duplex af relaying
  under residual self-interference,'' \emph{IEEE Wireless Communications
  Letters}, vol.~3, no.~2, pp. 233--236, Apr. 2014.

\bibitem{cheng2013twrfull}
X.~Cheng, B.~Yu, X.~Cheng, and L.~Yang, ``Two-way full-duplex
  amplify-and-forward relaying,'' in \emph{Proc. IEEE Military Commun. Conf.},
  Oct. 2013, pp. 1--6.

\bibitem{Tabataba12PA}
F.~S. Tabataba, P.~Sadeghi, C.~Hucher, and M.~R. Pakravan, ``Impact of channel
  estimation errors and power allocation on analog network coding and routing
  in two-way relaying,'' \emph{IEEE Transactions on Vehicular Technology},
  vol.~61, no.~7, pp. 3223--3239, Oct. 2012.

\bibitem{kim2013effects}
D.~Kim, H.~Ju, S.~Park, and D.~Hong, ``Effects of channel estimation error on
  full-duplex two-way networks,'' \emph{IEEE Transactions on Vehicular
  Technology}, vol.~62, no.~9, pp. 4666--4672, Oct. 2013.

\bibitem{zheng2015joint}
G.~Zheng, ``Joint beamforming optimization and power control for full-duplex
  {MIMO} two-way relay channel,'' \emph{IEEE Transactions on Signal
  Processing}, vol.~63, no.~3, pp. 555--566, Oct. 2015.

\bibitem{li2016fulltwr}
X.~Li, C.~Tepedelenlio\u{g}lu, and H.~\c{S}enol, ``Channel estimation for
  residual self-interference in full duplex amplify-and-forward two-way
  relays,'' \emph{IEEE Transactions on Wireless Communications}, vol.~16,
  no.~8, pp. 4970--4983, Oct. 2017.

\bibitem{Mohammadi16}
M.~Mohammadi, B.~K. Chalise, H.~A. Suraweera, C.~Zhong, G.~Zheng, and
  I.~Krikidis, ``Throughput analysis and optimization of wireless-powered
  multiple antenna full-duplex relay systems,'' \emph{IEEE Transactions on
  Communications}, vol.~64, no.~4, pp. 1769--1785, Oct. 2016.

\bibitem{Lemos15}
J.~S. Lemos, F.~A. Monteiro, I.~Sousa, and A.~Rodrigues, ``Full-duplex relaying
  in {MIMO-OFDM} frequency-selective channels with optimal adaptive
  filtering,'' in \emph{Proc. IEEE Global Conf. on Signal and Inf. Process.},
  Oct. 2015, pp. 1081--1085.

\bibitem{Cirik16}
A.~C. Cirik, M.~C. Filippou, and T.~Ratnarajaht, ``Transceiver design in
  full-duplex {MIMO} cognitive radios under channel uncertainties,'' \emph{IEEE
  Transactions on Cognitive Communications and Networking}, vol.~2, no.~1, pp.
  1--14, Oct. 2016.

\bibitem{Day12}
B.~P. Day, A.~R. Margetts, D.~W. Bliss, and P.~Schniter, ``Full-duplex {MIMO}
  relaying: Achievable rates under limited dynamic range,'' \emph{IEEE Journal
  on Selected Areas in Communications}, vol.~30, no.~8, pp. 1541--1553, Sep.
  2012.

\bibitem{Day12bi}
------, ``Full-duplex bidirectional {MIMO}: Achievable rates under limited
  dynamic range,'' \emph{IEEE Transactions on Signal Processing}, vol.~60,
  no.~7, pp. 3702--3713, Jul. 2012.

\bibitem{Tagh15}
O.~Taghizadeh, M.~Rothe, A.~C. Cirik, and R.~Mathar, ``Distortion-loop analysis
  for full-duplex amplify-and-forward relaying in cooperative multicast
  scenarios,'' in \emph{Proc. IEEE Int. Conf. Signal Process. and Commun.
  Syst.}, Oct. 2015, pp. 1--9.

\bibitem{Tagh16}
O.~Taghizadeh, T.~Yang, A.~C. Cirik, and R.~Mathar, ``Distortion-loop-aware
  amplify-and-forward full-duplex relaying with multiple antennas,'' in
  \emph{Proc. IEEE Int. Symp. Wireless Commun. Syst.}, Oct. 2016, pp. 54--58.

\bibitem{Ngo2014mimofull}
H.~Q. Ngo, H.~A. Suraweera, M.~Matthaiou, and E.~G. Larsson, ``Multipair
  full-duplex relaying with massive arrays and linear processing,'' \emph{IEEE
  Journal on Selected Areas in Communications}, vol.~32, no.~9, pp. 1721--1737,
  Oct. 2014.

\bibitem{Xiong16}
X.~Xiong, X.~Wang, T.~Riihonen, and X.~You, ``Channel estimation for
  full-duplex relay systems with large-scale antenna arrays,'' \emph{IEEE
  Transactions on Wireless Communications}, vol.~15, no.~10, pp. 6925--6938,
  Oct. 2016.

\bibitem{li2015maximum}
X.~Li and C.~Tepedelenlio\u{g}lu, ``Maximum likelihood channel estimation for
  residual self-interference cancellation in full duplex relay,'' in
  \emph{Proc. IEEE 49th Asilomar Conf. Signals, Syst. Comput.}, Nov. 2015, pp.
  807--811.

\bibitem{delayfilter}
\BIBentryALTinterwordspacing
J.~B. Calvert. Analog delay devices. [Online]. Available:
  \url{http://mysite.du.edu/~etuttle/electron/elect39.htm}
\BIBentrySTDinterwordspacing

\bibitem{riihonen2009power}
T.~Riihonen, S.~Werner, and R.~Wichman, ``Optimized gain control for
  single-frequency relaying with loop interference,'' \emph{IEEE Transactions
  on Wireless Communications}, vol.~8, no.~6, pp. 2801--2806, Oct. 2009.

\bibitem{MK1993}
S.~M. Kay, \emph{Fundamentals of Statistical Signal Processing: Estimation
  Theory}.\hskip 1em plus 0.5em minus 0.4em\relax Englewood Cliffs, NJ, USA:
  Prentice-Hall, 1993.

\bibitem{Kevin2012learning}
K.~P. Murphy, \emph{Machine Learning: A Probabilistic Perspective}.\hskip 1em
  plus 0.5em minus 0.4em\relax Cambridge, MA, USA: MIT Press, 2012.

\bibitem{Gray2006}
R.~M. Gray, \emph{Toeplitz and Circulant Matrices: A Review}.\hskip 1em plus
  0.5em minus 0.4em\relax LP Breda, The Netherlands: Now Publishers, Oct. 2006.

\bibitem{Goldsmithwireless}
A.~Goldsmith, \emph{Wireless Communications}.\hskip 1em plus 0.5em minus
  0.4em\relax New York, NY, USA: Cambridge University Press, 2005.

\bibitem{Oppenheimsignal}
A.~Oppenheim, A.~Willsky, and S.~Nawab, \emph{Signals and Systems}.\hskip 1em
  plus 0.5em minus 0.4em\relax New Jersey, NJ, USA: Prentice Hall, 1997.

\end{thebibliography}
\end{document}